\documentstyle[prd,aps,preprint,graphicx]{revtex}
\tighten
\newcommand{\half}{\frac{1}{2}}
\newcommand{\quart}{\frac{1}{4}}
\newcommand{\intvecx}{\int d^3 x\,}
\newcommand{\Tr}{\mbox{Tr}\,}
\newcommand{\dnot}{\partial_{0}}
\newcommand{\be}{\begin{equation}}
\newcommand{\ee}{\end{equation}}
\newcommand{\bea}{\begin{eqnarray}}
\newcommand{\eea}{\end{eqnarray}}
\newcommand{\veck}{{\bf k}}
\newcommand{\vecx}{{\bf x}}
\newcommand{\vecy}{{\bf y}}
\newcommand{\ah}{\hat{a}}
\newcommand{\ahd}{\hat{a}^{\dagger}}
\newcommand{\bh}{\hat{b}}
\newcommand{\bhd}{\hat{b}^{\dagger}}
\newcommand{\qhat}{\hat{q}}
\newcommand{\phat}{\hat{p}}
\newcommand{\al}{\alpha}
\newcommand{\bt}{\beta}

\newcommand{\dl}{\delta}
\newcommand{\ep}{\epsilon}

\newcommand{\et}{\eta}

\newcommand{\lm}{\lambda}
\newcommand{\rh}{\rho}
\newcommand{\sg}{\sigma}

\newcommand{\om}{\omega}

\newcommand{\Gm}{\Gamma}
\newcommand{\Lm}{\Lambda}
\newcommand{\Sg}{\Sigma}
\newcommand{\Dl}{\Delta}
\begin{document}
\draft
\title{Thermalization in a Hartree Ensemble Approximation
to Quantum Field Dynamics}
\author{Mischa~Sall\'e\thanks{email: msalle@science.uva.nl},
Jan Smit\thanks{email: jsmit@science.uva.nl} 
and
Jeroen C.~Vink\thanks{email: jcvink@science.uva.nl}}
\address{Institute for Theoretical Physics, University of Amsterdam\\
 Valckenierstraat 65, 1018 XE Amsterdam, the Netherlands}
\date{19 March 2001}
\maketitle
\begin{abstract}
For homogeneous initial conditions,
Hartree (gaussian) dynamical approximations are known to have problems with
thermalization, because of insufficient scattering.
We attempt to improve on this by writing
an arbitrary density matrix as a superposition of
gaussian pure states and applying the
Hartree approximation to each member of such an ensemble.
Particles can then scatter via their back-reaction on the
typically inhomogeneous mean fields.
Starting from initial states which are
far from equilibrium we numerically compute the time evolution of
particle distribution functions and observe that they indeed display
approximate thermalization on intermediate time scales
by approaching a Bose-Einstein form. However, for
very large times the distributions drift towards classical-like
equipartition.
\end{abstract}
\pacs{11.15.Ha, 11.10.Wx}
\section{Introduction}
\label{secintro}
Non-perturbative computations in quantum field theory in real time are
important for our understanding of
the physics of the early universe
as well as dynamics of heavy ion collisions.
Real-time simulations may also give us a new
handle on the difficult problem of computations at finite chemical potential, 
e.g.\ in QCD.
Incorporating finite density does not necessarily pose
extra problems of principle, so 
taking time averages in a thermalized
ergodic system will provide us with
microcanonical expectation values.

The classical approximation has given very useful
results for the sphaleron rate (see \cite{BoMoRu00} and \cite{SmTa99} 
for the status in three and one spatial dimensions), thermalization
after preheating \cite{afterpreh}, non-equilibrium electroweak baryogenesis
\cite{newewbary}, as well as for studies of
equilibration and thermalization \cite{Pa97,AaBoWe00a,AaBoWe00b}.
With the inclusion of fermions it has
given encouraging results for finite density simulations \cite{AaSm34}.
However, it suffers from Rayleigh-Jeans divergences. To some extent these
can be ameliorated in scalar field theories \cite{AaSm12}, but for gauge
theories the problems are more severe \cite{Ar97,Nautaea}.
Large $n$ approximations have also been used for initial value
problems, with $O(n)$-type models.
The leading order in $1/n$ has given useful results for the description of
preheating dynamics in the early universe (see e.g.~\cite{1oNph} and
references therein) and for
the possibly disoriented chiral condensate in heavy ion collisions 
\cite{1oNDCC}, 
but it is generally considered to contain
insufficient scattering for describing thermalization at larger
times. This will be improved in next order in $1/n$,
where scattering comes into play, but full implementation
in field theory is hard. Furthermore,  
within quantum mechanics one finds instabilities \cite{MiCoDa97,Miea00}, and
it has been argued that systematically correcting in $1/n$ does not prevent
the approximation to break down at
times of order $\sqrt{n}$ \cite{RyYa00}. On the other hand, 
Schwinger-Dyson-like
approaches including scattering diagrams appear to give more
favorable results \cite{MiCoDa00} and have been found to lead to
equilibration in field theory \cite{BeCo00}.

The leading order large $n$ equations for the $O(n)$ model are almost identical
to the Hartree approximation for the single component scalar
field, and so the latter approximation is also 
not considered to be able to describe
thermalization. Yet, in this paper 
we shall present evidence for approximate thermalization using Hartree dynamics
in a toy model, the $\varphi^4$ model in 1+1 dimensions. The crucial difference with
previous studies is that our system is allowed to be arbitrarily inhomogeneous.
This has the effect that particle-like excitations can scatter 
through the intermediary of a mean field fluctuating in time {\em and space}, 
which in turn is created by the particles. (We used `Hartree' rather than
`large $n$' to avoid problems with would-be Goldstone
bosons in 1+1 dimensions.)

The Hartree approximation describes the dynamics in terms of a mean field
and a two-point correlation function. It corresponds to a 
gaussian density matrix in field space, centered around the
mean field with a width given by the two-point function 
(see e.g.~\cite{Miea00}). 
The two-point function can be conveniently described in terms of a
complete set of mode functions.
For a homogeneous initial state the mean field is homogeneous 
and the mode functions can conveniently be taken in the form of plane waves
labeled by a wave vector $\veck$. 
Typically,
only mode functions in a narrow $|\veck|$-band get 
excited by the time-dependent homogeneous mean field, 
through parametric resonance or spinodal instability.
The system equilibrates but does not thermalize
in this approximation and particle distribution functions show resonance
peaks instead of approaching
the Bose-Einstein distribution (see for example \cite{Boyea96}). 

It is instructive to compare with the classical approximation.  
Simulations in this case indicate no problem of principle
with thermalization (see \ \cite{Pa97,AaBoWe00a,AaBoWe00b} 
for quantitative studies). 
Starting from an initial ensemble of classical field configurations
$\rh_c[\varphi,\pi,t_{\rm in}]$
(with canonical field variables $\varphi$ and $\pi$),
suitable observables are found to become distributed according
to the classical canonical distribution $\exp(-\bt H[\varphi,\pi])$.
This distribution will not be reached starting with strictly 
homogeneous realizations, because then the dynamics is that of a simple
system with only
two degrees of freedom, i.e.\ the spatially constant $\varphi$ and $\pi$.
As initial conditions aiming at thermalization
these are unsuitable realizations, even if
$\rh_c[\varphi,\pi,t_{\rm in}]$ is homogeneous. 
The phase space distribution $\rh_c[\varphi,\pi,t_{\rm in}]$ may be homogeneous,
but realizations $\varphi(\vecx,t_{\rm in})$, 
$\pi(\vecx,t_{\rm in})$ are typically inhomogeneous.
%
Viewing the Hartree approximation as a semiclassical improvement,
we may expect that thermalization will improve
if some analogies of classical realizations are used as initial states.

To implement the idea, we note that an arbitrary density operator
can be formally
written as a superposition of gaussian pure states:\footnote{Operators
are indicated with a caret.} 
\be
\hat{\rh} = \int [d\varphi\, d\pi]\, \rh_q[\varphi,\pi]\, 
|\varphi,\pi\rangle\langle\varphi,\pi|.
\label{rhtocoh}
\ee
Here 
the $|\varphi,\pi\rangle$ are coherent states centered around
$\varphi(\vecx)= \langle\varphi,\pi|\hat\varphi(\vecx)|\varphi,\pi\rangle$ and 
$\pi(\vecx)= \langle\varphi,\pi|\hat\pi(\vecx)|\varphi,\pi\rangle$, 
and $\rh_q[\varphi,\pi]$ is a 
functional representing the density operator $\hat\rh$.
We interpret the $|\varphi,\pi\rangle\langle\varphi,\pi|$ 
as `realizations' of $\hat\rh$.
The distribution $\rh_q[\varphi,\pi]$ can be 
quite singular for non-classical states, but for suitable semiclassical states
or thermal states 
it is positive and intuitively
attractive \cite{MaWo91}. We give a brief review in appendix A. 

A thermal state like $\exp[-\bt\hat H]$ 
cannot be approximated very well by a gaussian 
if there are nontrivial interactions.
For example, with a double well potential there are in general multiple 
peaks in the field distribution, while a gaussian has a 
single peak. 
But if in the decomposition (\ref{rhtocoh})
a gaussian state $|\varphi,\pi\rangle\langle\varphi,\pi|$ has a reasonable weight, 
we can take it as an initial state and use the Hartree approximation to compute
the time evolution. We can then
compute time averages (as long as the approximation is good), 
and finally sum over initial
states according to (\ref{rhtocoh}). 
Such a description is semiclassical in so far as the mean
field describes a near-classical path and $\rh_q[\varphi,\pi]$ 
is positive. 
But note that in the Hartree approximation the gaussian fluctuations 
(the modes comprising the two-point function
-- these are the `particle-like excitations' alluded to above)
influence the `classical' field, i.e.\ the mean field of the `realization'.

For thermal equilibrium the functional $\rh_q[\varphi,\pi]$ 
is time-independent but it 
is not known for interacting systems. If
the time evolution could be followed exactly, we would
be able to reconstruct
its microcanonical version, 
assuming the system is sufficiently strongly ergodic.
With exact dynamics we can imagine starting from 
some initial $\rh_q[\varphi,\pi]$ which is reasonably
close to the target distribution,
wait for equilibration and subsequently compute time averages
over an arbitrarily long time span. 
With only an approximation to the dynamics (Hartree) the distribution may
deteriorate after some time and we
may have to stop and start again.

Crucial questions are now, does the system equilibrate sufficiently in the 
Hartree approximation, such that results are insensitive
to reasonable choices of the initial $\rh_q[\varphi,\pi]$? 
Does it thermalize approximately, e.g.\ do 
one-particle distribution functions get the appropriate thermal forms? 
How long does it take for the approximation to break down? 
And 
if the answers to these questions are sufficiently favorable,
can we obtain a reasonable approximation to the target equilibrium 
distribution at intermediate times starting with a convenient initial one? 

We study these issues in a simple model, 1+1 dimensional $\varphi^4$ theory.
Sect.\ 2 introduces the model and the 
gaussian approximation. 
An effective hamiltonian $H_{\rm eff}$
describing the gaussian dynamics is introduced 
in Sect.\ 3. In Sect.\ 4 we discuss vacuum and thermal stationary
state solutions. We note one of the flaws of
the Hartree approximation, the fact that it predicts a first order
phase transition where there should only be a cross-over 
(in 3+1 D one also gets a first order transition \cite{RoMa96} instead of the
expected second order; the inconsistency problem with coupling constant 
renormalization \cite{RoMa96} is absent in 1+1 dimensions). 
Numerical results for the evolution from initial out-of-equilibrium
distributions are presented in Sec.~\ref{secnum}. 
We introduce a
one particle distribution function $n_k(t)$ and compare its time-dependent form
with the Bose-Einstein distribution.
In Sect.\ \ref{secdamp}
we study correlations in time of the zero momentum mode of the
mean field, 
use them for estimating damping times.
The results are discussed in Sect.\ \ref{secconcl}.
In appendix A we discuss the representation (\ref{rhtocoh}), 
in B classical equipartition according to $H_{\rm eff}$.

\section{Gaussian approximation}
\label{secgauss}
We start with the Heisenberg 
field equation for the quantum field\footnote{
In this section we assume 3+1 dimensions.}
at times $x^0>0$,
\be
(- \partial^2 + \mu^2)\hat\varphi(x) + \lm\hat\varphi(x)^3 = 0.
\label{opeq}
\ee
For exact evaluation we would
have to specify the infinite set of matrix elements of $\hat\varphi(\vecx,0)$ and
$\dnot\hat\varphi(\vecx,0)$ as initial conditions. In practise, of course,
less detail is needed.
Taking the expectation value in an initial state at time $x^0=0$ leads to
\bea
\langle\hat\varphi(x)\rangle &=& \varphi(x),
\label{vdef}\\
\langle T\hat\varphi(x_1)\hat\varphi(x_2)\rangle &=& \varphi(x_1)\varphi(x_2)  -iG(x_1,x_2),\\
\langle T\hat\varphi(x_1)\hat\varphi(x_2)\hat\varphi(x_3)\rangle &=& \varphi(x_1)\varphi(x_2)\varphi(x_3) 
-i\varphi(x_1) G(x_2,x_3) + \mbox{2 perm.}\nonumber\\&&
 + (-i)^2 G(x_1,x_2,x_3),\\
\langle T\hat\varphi(x_1) \cdots\hat\varphi(x_4)\rangle &=& \varphi(x_1)\cdots\varphi(x_4)
-i \varphi(x_1)\varphi(x_2) G(x_3,x_4) + \mbox{6 perm.}\nonumber\\&&
+ \varphi(x_1) (-i)^2 G(x_2,x_3,x_4) + \mbox{3 perm.}\nonumber\\&&
+ (-i)^2 G(x_1,x_2) G(x_3,x_4) + \mbox{2 perm.} \nonumber\\&&
+ (-i)^3 G(x_1,\ldots,x_4),
\eea
etc. Here $T$ denotes time ordering and
\be
\langle \hat\varphi(x_1) \cdots \hat\varphi(x_n)\rangle \equiv 
\Tr \hat\rh\, \hat\varphi(x_1) \cdots \hat\varphi(x_n),
\ee
with $\hat\rh$ the initial density operator;
$\varphi$ is the mean field (or classical field) and
the $G's$ are correlation functions (connected Green functions). 
Taking the expectation value of (\ref{opeq}) and neglecting the three point
correlation function $G(x,x,x)$ 
gives the approximate equation
\be
[-\partial^2 + \mu^2  + \lm\varphi(x)^2 - 3i\lm G(x,x)] \varphi(x) = 0.
\label{meanfeq}
\ee
To use it we need an equation for the two-point function. Such an equation
can be found by multiplying (\ref{opeq}) by $\hat\varphi(y)$ and 
taking again the expectation value in the initial state.
This leads to the approximate equation
\be
[-\partial^2 + \mu^2 + 3\lm\varphi(x)^2 - 3i\lm G(x,x)]\, G(x,y) = \dl^4 (x-y),
\label{Geq}
\ee
where we used the canonical commutation relations and dropped
the three and four-point correlation functions. 
We shall comment on their neglect at the end of this section.
Since only the two-point function appears, Eqs.\ (\ref{meanfeq},\ref{Geq}) 
are exact if
the hamiltonian and density matrix are approximated by gaussian forms.
Given the neglect of the higher correlation functions
the initial density matrix does not have to be gaussian {\em per se},
but its non-gaussianity does not enter in Eqs.\ (\ref{meanfeq},\ref{Geq}).
For clarity we shall now assume the bra-kets $\langle \cdots\rangle$
to refer to a {\em gaussian density operator} $\rh$. Later we will consider
non-gaussian operators by further averaging over initial conditions,
as in (\ref{rhtocoh}), 
which will be indicated by $\overline{\langle \cdots\rangle}$.

An intuitive as well as practical way for computing the two-point function 
is in terms of mode functions $f_{\al}(x)$. We write
\be
-iG(x,y) = \theta(x^0-y^0) C(x,y) + \theta(y^0-x^0) C(y,x).
\ee
such that 
\be
C(x,y) = \langle [\hat\varphi(x)-\varphi(x)][\hat\varphi(y)-\varphi(y)]\rangle.
\label{Cdef}
\ee
It follows from (\ref{Geq}) that $C(x,y)$
satisfies the homogeneous equation ($\dl^4(x-y) \to 0$), in the variable
$x$ as well as in $y$, as if $\hat\varphi(x) - \varphi(x)$ satisfies this equation.
We can now introduce mode functions
$f_{\al}(x)$ satisfying the homogeneous equation, 
\be
[-\partial^2 + \mu^2 + 3\lm\varphi(x)^2 + 3\lm C(x,x)]\, f_{\al}(x) = 0, 
\label{eomf}
\ee
($-iG(x,x)=C(x,x)$) and write
\be
\hat\varphi(x)\stackrel{\rm g.a.}{=}\varphi(x) + \sum_{\al}\left[ \bh_{\al} f_{\al}(x) 
+ \bhd_{\al} f_{\al}^*(x)\right].
\label{phiga}
\ee
where the $\bh_{\al}$ and $\bhd_{\al}$ are spacetime independent and
`g.a.' means `gaussian approximation'.
The wave equation (\ref{eomf})
for the $f_{\al}$ is of the Klein-Gordon type and we 
require the mode functions to be orthogonal and complete in the Klein-Gordon
sense,
\bea
\intvecx \left[f_{\al}^*(x) i\dnot f_{\bt}(x) 
- i\dnot f_{\al}^*(x) f_{\bt}(x)\right] 
&=& \dl_{\al\bt},
\label{orthog1}\\
\intvecx \left[f_{\al}(x) i\dnot f_{\bt}(x) 
- i\dnot f_{\al}(x) f_{\bt}(x)\right] 
&=& 0,
\label{orthog2}\\
\sum_{\al}\left[-if_{\al}(x) \dnot f_{\al}^*(y) + i f_{\al}^*(x)\dnot f_{\al}(y)
\right]_{x^0 = y^0}
&=& \dl^3(\vecx-\vecy),\\
\label{complete1}
\sum_{\al}\left[f_{\al}(x) f_{\al}^*(y) - f_{\al}^*(x) f_{\al}(y)
\right]_{x^0 = y^0}
&=& 0,\\
\sum_{\al}\left[\dnot f_{\al}(x) \dnot f_{\al}^*(y) - \dnot f_{\al}^*(x) 
\dnot f_{\al}(y)
\right]_{x^0 = y^0}
&=& 0.
\label{complete2}
\eea
The above orthogonality and completeness
relations are preserved by the equation of motion (\ref{eomf}) for the 
$f_{\al}$.  
The canonical commutation relations for $\hat\varphi$ and $\dnot\hat\varphi$
translate into 
\be
[\bh_{\al},\bhd_{\bt}] = \dl_{\al\bt},\;\;\;\;
[\bh_{\al},\bh_{\bt}] = [\bhd_{\al},\bhd_{\bt}] = 0.
\ee
The initial condition implies $\langle \bh_{\al}\rangle = 0$ and we have to 
specify $E_{\al\bt}\equiv\langle\bh_{\al}\bh_{\bt}\rangle$
and $N_{\al\bt}\equiv\langle\bhd_{\al}\bh_{\bt}\rangle$. The matrices
$N$ and $E$ are subject to constraints following from their definition
as expectation values of operators in Hilbert space. We shall assume that
a Bogoliubov transformation 
$\bh_{\al} \to \sum_{\bt} [A_{\al\bt} \bh_{\bt} 
+ B_{\al\bt} \bhd_{\bt}]$
can be made such that $E_{\al\bt}\to 0$ and 
$N_{\al\bt} \propto \dl_{\al\bt}$. This transformation produces new mode 
functions which are linear combinations of the $f$ and $f^*$. 
In the new basis we only have to specify as initial conditions
\be
\langle\bhd_{\al}\bh_{\bt}\rangle \equiv n^0_{\al}\,\dl_{\al\bt},
\;\;\;\;
n^0_{\al} \geq 0,
\label{n0in}
\ee
in terms of which
\be
C(x,y) = \sum_{\al} \left[(1+n^0_{\al}) f_{\al}(x) f_{\al}^*(y)
+ n^0_{\al} f_{\al}^*(x) f_{\al}(y)\right].
\label{Cf}
\ee

Eq.\ (\ref{phiga}) expresses the fact that in the gaussian
approximation the field 
$\hat\varphi'(x)\equiv\hat\varphi(x) - \varphi(x)$ is a generalized free field, 
i.e.\ its correlation functions are completely 
determined by the two-point function. Its linear field equation
(i.e.\ (\ref{eomf}) with $f_{\al} \to \hat\varphi'$) is equivalent to
the Heisenberg equations of motion of the effective gaussian
hamiltonian operator
\be
\hat H_{\rm g.a.} = \intvecx \left[
\half\hat\pi^{\prime 2} +
\half(\nabla\hat\varphi')^2 + \half m_{\rm eff}^2 \hat\varphi^{\prime 2} + 
\ep_{\rm eff} \right],
\ee
where the spacetime dependent effective mass $m^2_{\rm eff}$ 
is given by
\be
m_{\rm eff}^2(x) = 3\lm\varphi(x)^2 + 3\lm C(x,x).
\label{meffdef}
\ee
We also introduced an effective c-number
energy density $\ep_{\rm eff}$, which is determined 
by requiring $\langle \hat H_{\rm g.a.} \rangle = \langle \hat H\rangle$:
\be
\ep_{\rm eff}(x) = \half\pi(x)^2 + \half[\nabla\varphi(x)]^2 
+ \half\mu^2\varphi(x)^2 + \quart \lm\varphi(x)^4 
- \frac{3}{4}\lm C(x,x)^2.
\ee

Summarizing, the gaussian
approximation consists of 
eqs.\ (\ref{meanfeq}), (\ref{eomf}), (\ref{n0in}) and (\ref{Cf}),
together with the orthogonality and completeness conditions 
(\ref{orthog1})-(\ref{complete2}) for the mode
functions and
some initial condition 
for the mean field and mode functions.

The gaussian approximation can be justified in the limit of large $n$
for the $O(n)$ model. 
The resulting field equations are very similar: we only need to
make the replacement $3\to 1$ in eqs.\ (\ref{meanfeq}) and (\ref{eomf}).

\begin{figure}[]
\includegraphics[width=\textwidth]{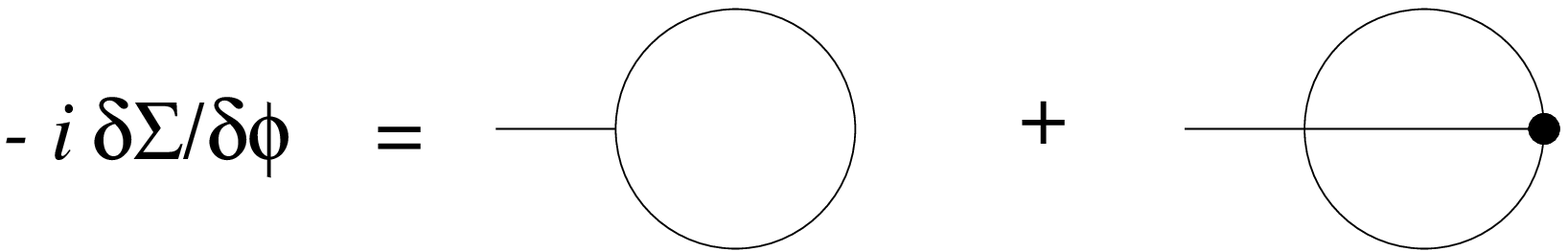}
\caption{Diagrammatic illustration of $\dl\Sg/\dl\varphi$, with 
$\Sg$ the selfenergy functional defined by $\Gm = S - \Sg$. The lines and
full dots represent the exact propagators (correlation functions) and 
vertex functions, the other vertices represent the bare vertex functions
as given by the classical action $S$.
} \label{f1}
\end{figure}

\begin{figure}[]
\includegraphics[width=\textwidth]{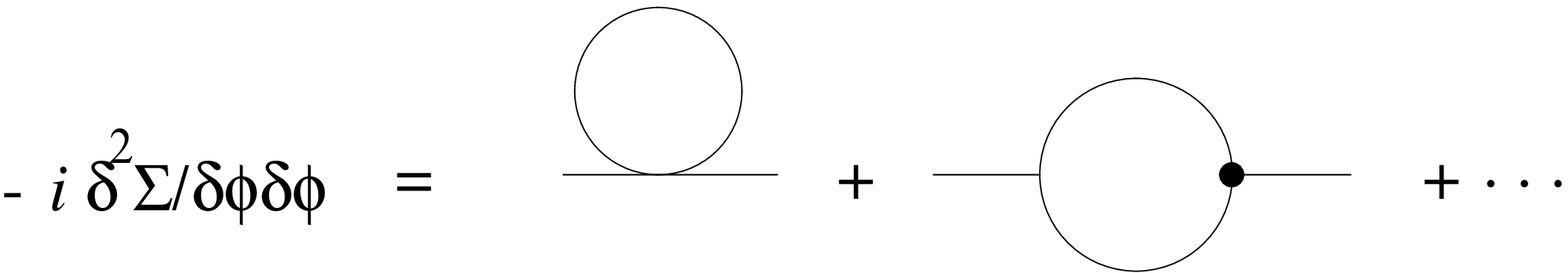}

\caption{Diagrams for the selfenergy part of the inverse correlation
function $G^{-1} = -\dl^2 S/\dl\varphi\,\dl\varphi + \dl^2\Sg/\dl\varphi\,\dl\varphi$.
The $\cdots$ represent the two-loop diagrams obtained by differentiating
the diagrams in Fig.~1.
} \label{f2}
\end{figure}

The above derivation in terms of
the Heisenberg equations of motion can be put into the systematic
framework of the Dyson-Schwinger hierarchy. These equations follow
from functionally differentiating an exact
equation of motion $\dl\Gm/\dl\varphi = -J$
with respect to $J$ and setting $J=0$ afterwards.
Here $\Gm$ is the effective action 
(with time integration along the usual Keldysh-Schwinger contour) and $J$ 
an external source. 
We shall not go into details here, but just
comment on the systematics, using diagrams
(for a derivation, see e.g.~\cite{DeWi64}). 
Fig.~\ref{f1}
illustrates the exact equation for the mean field. The gaussian
approximation (\ref{meanfeq}) is obtained by dropping the two-loop 
diagram. By differentiating the diagrams in Fig.~\ref{f1} we
get the exact equation for the two-point correlation function
illustrated in Fig.\ \ref{f2}. 
The gaussian approximation (\ref{Geq}) can be obtained from this by: a)
dropping the two-loop contributions and b) dropping the 
second one-loop diagram. The neglect of the two-loop terms 
may be reasonable at weak coupling, and even the second
approximation may be justifiable if the product of the three
point couplings (one bare, the other dressed) is substantially smaller
than the (bare) four point coupling in the first one-loop diagram.
However, since the bare three point vertex 
$\dl^3 S/\dl\varphi^3 \propto \lm\varphi$ we
see that this is not likely if $\varphi = O(\lm^{-1/2})$ or larger.
Especially this second approximation b)
is worrisome, because on iteration of the integral equations we
would not get correctly all one-loop diagrams. 
It is also known
that the approximation does not give exact Goldstone bosons where
one expects them, because the phase transition is incorrectly predicted to
be first order, instead of second order (in 3+1 D) or a cross
over (1+1 D). There is a problem with renormalization in 3+1
dimensions \cite{RoMa96} (but not in 1+1 D). 

It will depend on the circumstances if these troublesome features of
the Hartree approximation
are numerically important.

\section{Effective hamiltonian and conserved charges}
\label{seceffh}
The equations of the gaussian approximation derived in section \ref{secgauss}
are local in time 
and they may be derived from
a conserved effective hamiltonian. We shall present it here
and exhibit its symmetries and accompanying conserved charges.
We write
\bea
f_{\al}(x) &=&\frac{1}{\sqrt{2}}\left[f_{\al 1}(x) -if_{\al 2}(x)\right],\\
\xi_{\al a}(x) &=& 
\left(\half + n^0_{\al}\right)^{1/2}
f_{\al a}(x),\;\;\;\; a=1,2.\\
\et_{\al a}(x) &=& \dnot \xi_{\al a}(x),\;\;\;\; \pi(x) = \dnot \varphi(x).
\label{xiintro}
\eea
In terms of the real canonical
variables $\varphi$, $\pi$, $\xi_{\al a}$ and $\et_{\al a}$
the effective hamiltonian takes the form
\bea
H_{\rm eff} &=& \int d^3 x\, \left[
\half \left(\pi^2 + \et^2 + (\nabla\varphi)^2 + (\nabla\xi)^2
\right)\right.\nonumber\\
&& \left.\mbox{}
+\half\mu^2\left(\varphi^2 + \xi^2\right)
+ \quart\lm\left(\varphi^4 + 6\varphi^2\xi^2 + 3 (\xi^2)^2\right)
\right],
\label{heff}
\eea
where
\be
\xi^2 = \sum_{\al}\left(\xi_{\al 1}^2 + \xi_{\al 2}^2\right),
\;\;\;
(\nabla\xi)^2 = 
\sum_{\al}\left[(\nabla\xi_{\al 1})^2 + (\nabla\xi_{\al 2})^2\right],
\;\;\;
\et^2 = \sum_{\al}\left(\et_{\al 1}^2 + \et_{\al 2}^2\right).
\ee
It is easy to check that the mean field equation (\ref{meanfeq}) 
and the mode equations (\ref{eomf}) are equivalent to the Hamilton equations
\be
\dnot \varphi = \pi,\;\;\;
\dnot \pi = -\frac{\dl H_{\rm eff}}{\dl \varphi},\;\;\;\;
\dnot \xi_{\al a} = \et_{\al a},\;\;\;\;
\dnot \et_{\al a} = - \frac{\dl H_{\rm eff}}{\dl \xi_{\al a}}.
\ee
It is also straightforward to show that $H_{\rm eff}$ is just the expectation
value of the quantum hamiltonian $\hat H(t)$ upon inserting the gaussian
approximation (\ref{phiga}), 
\be
H_{\rm eff} = \langle \hat H\rangle.
\ee

The effective
hamiltonian has evidently a large symmetry corresponding to rotations
of the infinite dimensional vectors $\xi_{\al a}$ and $\et_{\al a}$.
For definiteness, let us assume a regularization of the field theory such that 
there are $M$ modes, $\al = 1,\ldots,M$
(e.g.\ on an $N^3$ periodic lattice $M=N^3$).
Then the effective hamiltonian has $O(2M)$ symmetry, implying $M(2M-1)$
conserved generalized angular momenta of the general form
\be
L_{\al a,\bt b} = \intvecx \left(\xi_{\al a}\et_{\bt b} 
- \xi_{\bt b} \et_{\al a}\right),\;\;\;(\al,a)\neq(\bt,b).
\label{lang}
\ee
Recalling the orthonormality relations for the mode functions (\ref{orthog1}),
(\ref{orthog2}) we see that the conserved quantities are given in terms
of the initial conditions as
\be
L_{\al 1,\al 2} = \half + n^0_{\al},
\label{Lconstr}
\ee
with all others vanishing. 

It is interesting to compare with the effective hamiltonian corresponding to
the large $n$ limit of the $O(n)$ model \cite{Cooea97}, 
which may be obtained from
$H_{\rm eff}$ above by the replacement $3 \to 1$ (and $6\to 2$). This has
the effect of producing the combination $\lm(\varphi^2 + \xi^2)^2$, so
the symmetry enlarges to $O(2M+1)$. The additional $2M$ conserved generalized 
angular momenta depend on the initial conditions for $\varphi$ and $\pi$.\footnote{
In \cite{Cooea97} the effective hamiltonian for the homogeneous system
was expressed in terms of the radial variable
$\xi_{\al} = \sqrt{\xi_{\al 1}^2 + \xi_{\al 2}^2}$ (modulo a factor of two),
and the rotational symmetries mixing $\xi_{\al 1}$ and $\xi_{\al 2}$
are then absent. 
However, the corresponding equations
of motion then suffer from numerical complications due to the angular
momentum barriers.}

\section{Equilibrium states}
\label{secground}

In a first exploration of the system and of the gaussian
approximation we study equilibrium states, i.e.\ stationary states with
 maximum entropy.
This will give information on the phase
structure and quasiparticle excitations as a function of temperature.
From now on we specialize to 1+1 dimensions, $x^{\mu}\to (x,t)$, and assume the 
system to have `volume' $L$ with periodic boundary conditions. 
The coupling $\lm$ needs no renormalization while the bare mass parameter 
$\mu^2$ is only logarithmically divergent with the implicit cutoff.
 
We assume the equilibrium states to be homogeneous and time-independent,
i.e.\ $\varphi(x,t) = v$ and $C(x,t;y,t) = C(x-y,0;0,0)$. Also the various
time derivatives of $C$ evaluated at equal times are assumed to be 
time-independent.
We shall seek solutions of the form (\ref{Cf}) in which the mode 
functions are plane waves,
\be
\varphi(x,t) = v,\;\;\;\;
f_k(x,t) = \frac{e^{ikx-i\om_k t}}{\sqrt{2\om_k L}}.
\label{modevac}
\ee
Here the label $\al$ is the wave number $k$ and
we write $n_k$ for the corresponding (time independent) occupation numbers.
With this ansatz the equations for the mean field and mode functions 
reduce to
\bea
(\mu^2 + 3\lm C + \lm v^2) v &=& 0,
\label{meq2}
\\
-\om_k^2 + k^2 + \mu^2 + 3\lm C + 3\lm v^2 &=& 0,
\eea
where $C=C(x,t;x,t)$ is time-independent. 
In the infinite volume limit it is given by
\be
C = \int \frac{dk}{2\pi}\,\left(n_k + \half\right)\, 
\frac{1}{\om_k}. 
\ee
It follows that
\be
\om_k^2 = m^2 + k^2, \;\;\;\;
m^2 = \mu^2 + 3\lm C + 3\lm v^2.
\label{meq1}
\ee
To determine the $n_k$ we 
maximize the entropy $S$ subject to the constraint of fixed energy 
$U\equiv H_{\rm eff}=E$,
i.e.\ maximize $S + \bt (E-U)$, with Lagrange multiplyer $\bt$. We shall write
these equations in terms of the densities $s=S/L$, $u=U/L$, $\ep = E/L$
with $L\to\infty$.
The (unrenormalized) energy density $u$ is given by
\be
u = \frac{H_{\rm eff}}{L} = 
\half \mu^2 v^2 + \quart \lm v^4 
+ \int \frac{dk}{2\pi}\,
\left(n_k+\half\right) \frac{\om_k^2 + k^2 + \mu^2 + 3\lm v^2}{2\om_k} 
+ \frac{3}{4}\,\lm C^2,
\label{u}
\ee
and for our gaussian density operator, $s$ can be written as
\be
s = -\frac{1}{L}\Tr \rh\log\rh = 
\int \frac{dk}{2\pi}\,
\left[\left(n_k+1\right)\log\left(n_k+1\right)-n_k\log n_k\right].
\label{qentrop}
\ee
The 
maximization equations read
\be
0=\frac{\dl [s+ \bt(\ep-u)]}{\dl n_k} = 
\log\left(\frac{n_k +1}{n_k}\right) - \bt \om_k,
\;\;\;\;
u=\ep,
\ee
with the solution 
\be
n_k = \frac{1}{e^{\bt\om_k}-1}
\label{BEform}
\ee
and $\bt$ such that $u=\ep$.
So we found equilibrium states of the Hartree evolution corresponding to
the Bose-Einstein distribution with temperature $T=\bt^{-1}$. 
All effects of the interaction are buried in the temperature dependent
mass $m$ introduced in (\ref{meq1}).

For simplicity of discussion, let us next
use a simple momentum cutoff $|k|<\Lm$
and define a renormalized mass parameter $\mu_r^2$ by 
\be
\mu_r^2 = \mu^2 + \frac{3\lm}{4\pi}\log\frac{4\Lm^2}{\lm}.
\ee
Then (\ref{meq1}) takes the renormalized form
\be
m^2 = \mu_r^2 + \frac{3\lm}{4\pi}\log\frac{\lm}{m^2} + 
3\lm\int_0^{\infty} \frac{dk}{\pi}\, \frac{1}{\sqrt{m^2 + k^2}}\,
\frac{1}{e^{\sqrt{m^2 + k^2}/T}-1}
+ 3\lm v^2.
\label{mvT}
\ee

\begin{figure}[]
\begin{center}
\includegraphics[width=\textwidth]{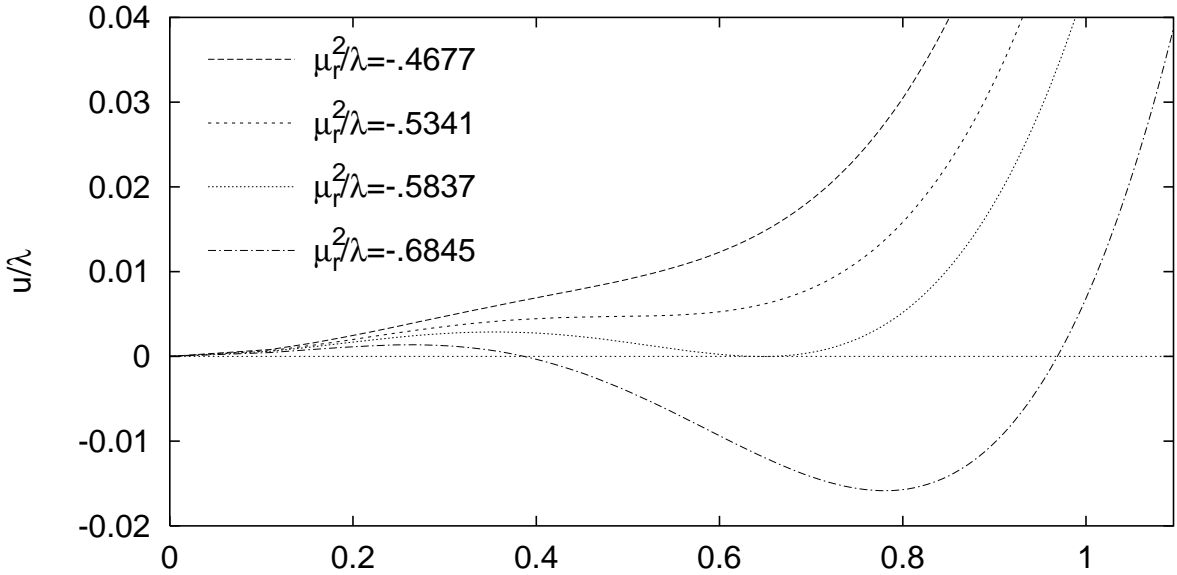}
\caption{Zero temperature effective potential $u/\lm = H_{\rm
eff}/L\lm$ versus $\varphi$ for various values of $\mu_r^2/\lm$. The
potential is normalized to zero at $\varphi = 0$. \label{figeffpotT0} }
\end{center}
\end{figure}

At zero temperature the equilibrium state is 
the vacuum. For $v=0$ there is one solution $m^2$ for every 
$\mu_r^2 \in (-\infty,\infty)$. For nonzero $v$ we get with (\ref{meq2})
the relations
\be
m^2 = 2\lm v^2,
\;\;\;\;
\mu_r^2 = -\half m^2 + \frac{3\lm}{4\pi} \log\frac{m^2}{\lm} .
\ee
There turn out to be {\em two} solutions, provided 
$\mu_r^2/\lm < (3/4\pi)[-1+\log(3/2\pi)]\approx -0.415$, otherwise none. 
To determine the true ground state we plot in Fig.~\ref{figeffpotT0}
the effective potential $u$ as a function 
$\varphi$ (i.e.\ $m^2$ is the solution of (\ref{mvT}) with $v\to\varphi$ at $T=0$), 
for
various $\mu_r$. The plot shows that there is a first order phase transition
as a function of $\mu_r^2$, instead of the expected second order transition
for a model in the universality class of the Ising model. This 
mis-representation of the phase transition is a well-known
artefact of the gaussian approximation (see, e.g.\ \cite{RoMa96}). 

Note that the 2nd order transition
would occur at strong coupling $\lm/m^2\to\infty$, where the 
gaussian approximation is suspect. In fact, the two masses
at the transition also imply strong coupling: they are given by 
$\lm/m^2 \approx 10$, for $\varphi = 0$ and 
$\lm/m^2 \approx 1.2$ for $\varphi = v_c \approx 0.65$.
To avoid fake first order effects we should evidently choose parameters 
away from the transition region. 
For this paper we mostly used
$\lm/m^2 = 1/12$ for which there is only one ground state
at $v^2 = 6$, well away from $v_c^2\approx 0.65$.

Having determined the groundstate
we define the renormalized energy $H_{\rm eff,r}$
by subtracting from $H_{\rm eff}$ 
its value in the ground state, such that the vacuum energy is zero.
It can be instructive to split the total energy into a classical
(gaussian mean field) part and a mode energy,
$H_{\rm eff,r} = H_{\rm clas} + H_{\rm modes}$, where we define the classical
part as
\bea
H_{\rm clas} &=&\int dx\, \left[\half\pi^2 + \half(\nabla\varphi)^2
+ V_{\rm clas}(\varphi)\right],
\label{Eclas}\\
V_{\rm clas}(\varphi)&=& \half m^2\varphi^2 + \quart \lm \varphi^4,
\;\;\;\;
v=0,\\
&=& \quart\lm\left(\varphi^2 - v^2\right)^2,
\;\;\;\; v\neq 0,
\eea
where $m^2$ and $v^2$  are the vacuum values ($T=0$).
 
\begin{figure}[]
\begin{center}
\includegraphics[width=\textwidth]{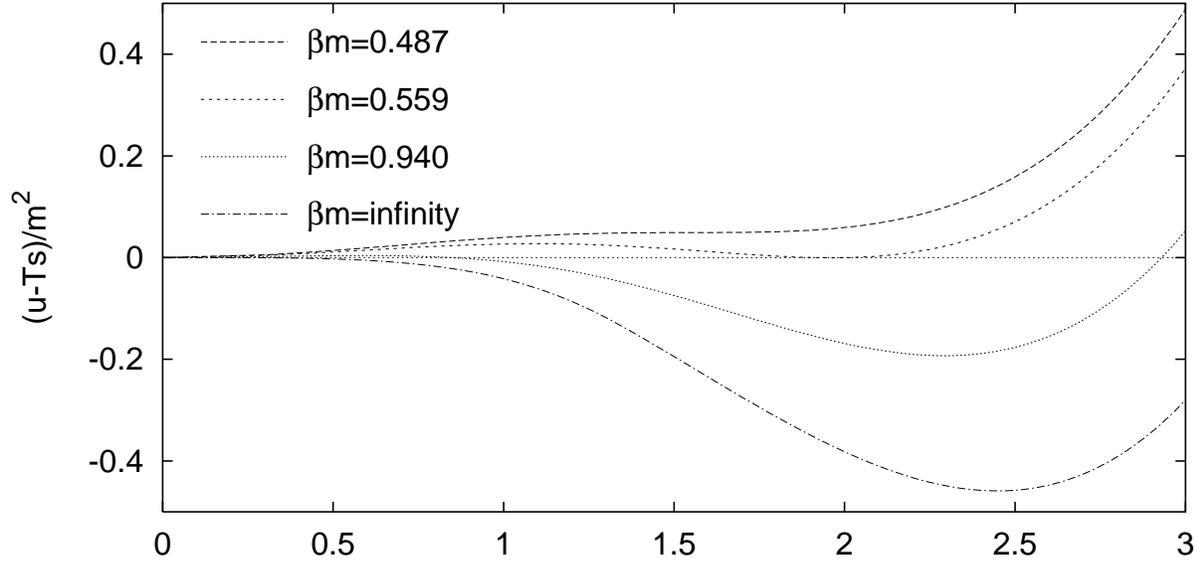}
\caption{Finite temperature effective potential $f/\lm = (u-Ts)/\lm$ 
versus $\varphi$ for various values of $\bt m(\varphi_c,0)$. The
potential is again normalized to zero at $\varphi = 0$. \label{figeffpotT} }
\end{center}
\end{figure}

Consider now starting in the broken symmetry phase $v\neq 0$
at zero temperature and raising the temperature. 
In 1+1 dimensions there should be only a cross over 
and not a true phase transition.
Fig.\ \ref{figeffpotT} shows the finite temperature effective potential
(free energy density)
\be
f(\varphi) = u(\varphi) - T s(\varphi),
\label{eqeffpotT}
\ee
using the temperature $T$ as independent
variable instead of the energy density $\ep$.
Now $m^2= m^2(\varphi,T)$ is the solution of (\ref{mvT}),
$v \to \varphi$, at finite $T$. The parameters were chosen such that
$v^2 = m^2(v,0)/2\lm = 6$ at $T=0$.
We see again a fake first order transition, at
$T_c \approx 1.79\, m(v,0)$, with $v_c = 1.96$. 
Its latent heat $\ell$ and surface tension $\sg$ are
given by $\ell = \Dl u = 0.39\, m(v,0)^2$, 
$\sg = \int_0^{v_c} d\varphi\, \sqrt{2 f(\varphi)} = 0.295\, m(v,0)$.
These are not particularly small values and we may not argue that the
effects of the first order transition will be negligible under generic
circumstances. However, the critical size of a nucleating
bubble is zero in 1+1 dimensions, so the bubble nucleation rate 
is not suppressed ($\propto \exp(-2\sg/T_c) \approx \exp(-0.17)$) 
and supercooling will not be strong.


We end this section with some cautionary remarks. 
First, the fact that the equilibrium
correlation function $C(x,y)$ has the free form (i.e.\ eq.\ 
(\ref{Cequill}) below with $n_k$ given by the Bose-Einstein form (\ref{BEform}))
for any coupling strength is a result of
the gaussian approximation.  The exact correlation function will have a more 
complicated form, although the corrections are expected to be small
at weak coupling. We will check this by a Monte Carlo computation in
a separate publication \cite{SaSmVi00c}.

Second,
it is not clear that the finite temperature 
equilibrium state found above will actually be 
approached at very large times. 
Any set of numbers $n_k$ in conjunction with
Eqs.\ (\ref{modevac})--(\ref{u}) gives a stationary solution to the
Hartree equations. Our derivation of 
the Bose-Einstein form for $n_k$ used the standard form (\ref{qentrop}) 
for the entropy, but we have not shown that this entropy 
is a large time result of the dynamics. Of course, this would be trivially
the case if we choose the initial ocupation numbers $n^0_{\al}=n_k$. But 
for a generic gaussian initial state the correlation function may still
approach a fixed point of the form just discussed ($t\approx t'$),
\bea
C(x,t;x',t') &=& \sum_{\al} \left[(1+n^0_{\al}) f_{\al}(x,t) f_{\al}^*(x',t')
+ n^0_{\al} f_{\al}^*(x,t) f_{\al}(x',t')\right]
\nonumber\\
&\to& \int \frac{dk}{2\pi}\, \left[\frac{1+n_k}{2\om_k}\,
e^{ik(x-x')-i\om_k(t-t')} + \frac{n_k}{2\om_k}\,
e^{-ik(x-x')+i\om_k(t-t')}\right],
\label{Cequill}
\eea
where the $n_k$ are expected to correspond to maximum entropy in relation
to the dynamics.  Since the Hartree dynamics in terms of 
$H_{\rm eff}$ is classical we may expect this entropy
to take a classical form, which would lead to
\be
n_k = \frac{T}{\om_k}.
\label{nclas}
\ee
Matters are complicated by the presence
of the infinitely many conserved charges (\ref{Lconstr}), which are
determined by the initial conditions. Note that without these constraints
one would expect $n_k+1/2 =T/\om_k$, instead of (\ref{nclas}), which makes
a big difference because equipartition suggests low 
$T = O(\ep/\Lm)$ 
and therefore {\em small} $n_k$.
We elaborate on this in appendix B.

To study such matters numerically we now first
introduce a coarse graining of the correlation function and define a
corresponding time dependent distribution function $n_k(t)$.

\section{Coarse grained particle numbers}
\label{sectcoarse}

The mode functions may be interpreted as representing particles
which interact through the mean field. This is
similar to electrons scattering off each other in 
classical electrodynamics, albeit that here
the `particles' are treated quantum mechanically and their interaction
is short ranged.  
Intuitively, such an interpretation supposes that the particles are
localized, with a correspondingly fluctuating (and hence inhomogeneous)
mean field taking the role of a classical field.

Within such a picture one expects the system to thermalize approximately.
We would like such thermalization to be quantal,
e.g.\ with particle distribution functions which are of the Bose-Einstein
type. However, the fact that our equations of motion have the form of
classical Hamilton equations in terms of $H_{\rm eff}$ suggests otherwise,
namely a distribution approaching a classical Boltzmann form
$\exp(-\bt H_{\rm eff})$, subject to the constraints set by the large number
of conserved charges (\ref{lang}). But this may take a very long time.
In any case, one way to test the gaussian approximation is to study its 
thermalization properties.

This we do by looking at 
equal time correlation functions, coarse grained by
averaging over a spacetime region.
Assuming the system is weakly coupled we can compare such averages with
a free field form in terms of quasiparticles with effective masses.  
If the system equilibrates locally in a quantum way,
then the quasiparticle distribution $n_k$ should approach 
the Bose-Einstein form. 
We define the correlation functions 
\bea
S(x,y,t) &=& \overline{\langle\hat\varphi(x,t)\hat\varphi(y,t)\rangle}
- \overline{\langle\hat\varphi(x,t)\rangle}\;\overline{\langle\hat\varphi(y,t)\rangle},\\
T(x,y,t) &=& \frac{1}{2}\overline{
\langle\left[\hat\varphi(x,t)\hat\pi(y,t) + \hat\pi(y,t)\hat\varphi(x,t)\right]\rangle}
-\overline{\langle\hat\varphi(x,t)\rangle}\;\overline{\langle\hat\pi(y,t)\rangle},\\
U(x,y,t) &=& \overline{\langle\hat\pi(x,t)\hat\pi(y,t)\rangle}
-\overline{\langle\hat\pi(x,t)\rangle}\;\overline{\langle\hat\pi(y,t)\rangle},
\eea
where the overbar denotes the spacetime averaging as well as a
possible average
over initial conditions as in (\ref{rhtocoh}). 
Using (\ref{vdef}) and (\ref{Cdef})
we can express these quantities in terms of a `classical' (mean field)
and a `quantum' contribution, 
\bea
S(x,y,t) &=& S^c(x,y,t) + S^q(x,y,t),
\label{cq1}\\
S^c(x,y,t) &=& \overline{\varphi(x,t)\varphi(y,t)} 
- \overline{\varphi(x,t)}\; \overline{\varphi(y,t)},\\
S^q(x,y,t) &=& \overline{C(x,t;y,t)},
\label{cq2}
\eea
etc. Note that $S^c \to 0$ in case of averaging over initial conditions
and/or spacetime. 

For simplicity the spatial average is performed over all of space.
For example,
\be
\overline{\langle\hat\varphi(x,t)\hat\varphi(y,t)\rangle}
=\frac{1}{L\Dl}\int_{t-\Dl/2}^{t+\Dl/2} dt'\, \int_0^L dz\,
\langle\hat\varphi(x+z,t')\hat\varphi(y+z,t')\rangle. 
\ee
Because of the periodic boundary conditions $S$, $T$ and $U$ depend only
on the difference between $x$ and $y$. Taking the Fourier transform
\be
S_k(t) = \frac{1}{L}\int_0^L dx\, dy\, e^{-ik(x-y)}\, S(x,y,t),
\;\;\;\;
k=(0, \pm 1, \pm2, \cdots)\frac{2\pi}{L},
\ee
and similarly for $T$ and $U$, it is easy to see that $S$ and $U$ are symmetric
and positive, i.e.\
\be
S_k(t) = S_{-k}(t) \geq 0,\;\;\;\; U_k(t) = U_{-k}(t) \geq 0,
\ee
while $T_k$ enjoys no such properties. For a free field with 
average occupation numbers $\langle \ahd_k\ah_k\rangle = n_k$ and frequencies
$\om_k$ the correlators are given by
$S_k = (n_k + n_{-k} + 1)/2\om_k$, $T_k = (n_k - n_{-k})/2$ and
$U_k = S_k \om_k^2$. Note that in this case $T$ is antisymmetric.
We now {\em define} $\om_k(t)$ and $n_k(t)$ for the interacting case by
\bea
\label{nomdef1}
n_k(t) &=& n_k^s(t) + n_k^a(t),\;\;\;\;
n_k^s(t) = n_{-k}^s(t),\;\;\;
n_k^a(t) = -n_{-k}^a(t),\\
\label{nomdef2}
S_k(t) &=& \left[n_k^s(t) + \half\right]\frac{1}{\om_k(t)},\\
\label{nomdef3}
T^a_k(t)&=& \half\left[T_k(t) - T_{-k}(t)\right] = n_k^a(t),\\
\label{nomdef4}
U_k(t) &=& \left[n_k^s(t) + \half\right]\,\om_k(t).
\eea
These equations can be easily solved in terms of $\om_k$ and $n_k$:
$\om_k= \om_{-k} = \sqrt{U_k/S_k}$, $n_k^s = \om_k S_k-1/2$ and
$n_k$ follows by adding $T_k^a$. 

There is a more direct interpretation of these formulas in terms
of the expectation value of a number operator $\ahd_k\ah_k$.
Suppose we define time dependent
creation and annihilation operators as
\be
\ah_k(t) = \frac{1}{\sqrt{2\om_k(t) L}}\int_0^L dx\, e^{-ikx}\, 
[\om_k(t)\hat\varphi(x,t) + i \hat\pi(x,t)],
\;\;\;\;
\ahd_k(t) = \left(\ah_k(t)\right)^{\dagger}.
\label{cadef}
\ee
Then 
\be
\overline{\langle \ahd_k(t)\ah_k(t)\rangle} = n_k(t).
\ee
The problem with starting with (\ref{cadef}) is that one does not
know a priori how to choose the $\om_k(t)$. This is especially so if
some of the effective squared 
frequencies $\mu^2 + 3\lm\varphi^2 + 3\lm C$
in the equations for the mode functions turn  
negative. The line of reasoning leading to 
(\ref{nomdef1}) -- (\ref{nomdef4}) solves this problem, but we should
keep in mind that this is by brute force, which can be misleading in
extreme situations, e.g.\ when the spectral function is not dominated by
a sufficiently narrow quasiparticle bump.

%
\section{Numerical results}
\label{secnum}
We now describe some simulations used for obtaining the particle numbers
$n_k(t)$. The mass and coupling parameters were chosen such that the system
at zero temperature is in the  `broken symmetry phase'.
The coupling was weak, $v^2 = m^2/2\lm = 6$. Here and
in the following $m$ is the mass of the particles at zero temperature.

The system is discretized on a space-time lattice with 
spatial (temporal) lattice distance $a$ ($a_0$), with
$a_0/a=0.1$. The number of spatial lattice sites, equal to the number of
independent complex mode functions, 
will be denotes with $N = L/a$.  The discretized lagrangian gives
rise to second order difference equations, with a time evolution which is 
equivalent to a first order leapfrog algorithm for 
$\pi_x(t)\equiv [\varphi_x(t+a_0)-\varphi_x(t)]/a_0$ and $\varphi_x(t)$.

The initialization is similar to that used in \cite{Pa97,AaBoWe00a},
\begin{equation}
\varphi_x = v, \qquad \pi_x = A m \sum_{j=1}^{j_{\rm max}} \cos(2 \pi j x/L -
\psi_j),
\end{equation}
with random phases $\psi_j$ uniformly distributed in $[0,2\pi)$. 
The modes 
are initialized with the equilibrium form at zero temperature: the
$n^0_k$ are all zero and the modes $f_k(x,0)$, $\dot{f}_k(x,0)$ are given
by the plane waves
(\ref{modevac}) and its time derivative at $t=0$, with 
$\om_k^2= k^2 + m^2$.
The density operator is thus a superposition of coherent pure
states as in (\ref{rhtocoh}).

We now describe a simulation for which 
$\lm/m^2 = 1/12$, $N=256$, $mL = 32$, 
$j_{\rm max} = 4$, $A=1/\sqrt{2}$,
such that 
the energy density is given by 
$E/Lm^2 = A^2 j_{\rm max}/4= 0.5$.
A Bose-Einstein distribution describing particles with such an
energy density would have a temperature $T/m \approx 1.08$, well below the
phase transition at $T/m \approx 1.8$, 
as calculated with the finite temperature effective potential. 
We also chose these parameters so that the system may end up in a 
low temperature quantum 
regime and not in a classical regime with $T/m\gg 1$.
A boring consequence was that the volume averaged mean field typically just 
oscillated around one of the two
minima, we did not encounter an initial condition for which
it crossed the barrier after $tm > 50$.

\begin{figure}[]
\includegraphics[width=\textwidth]{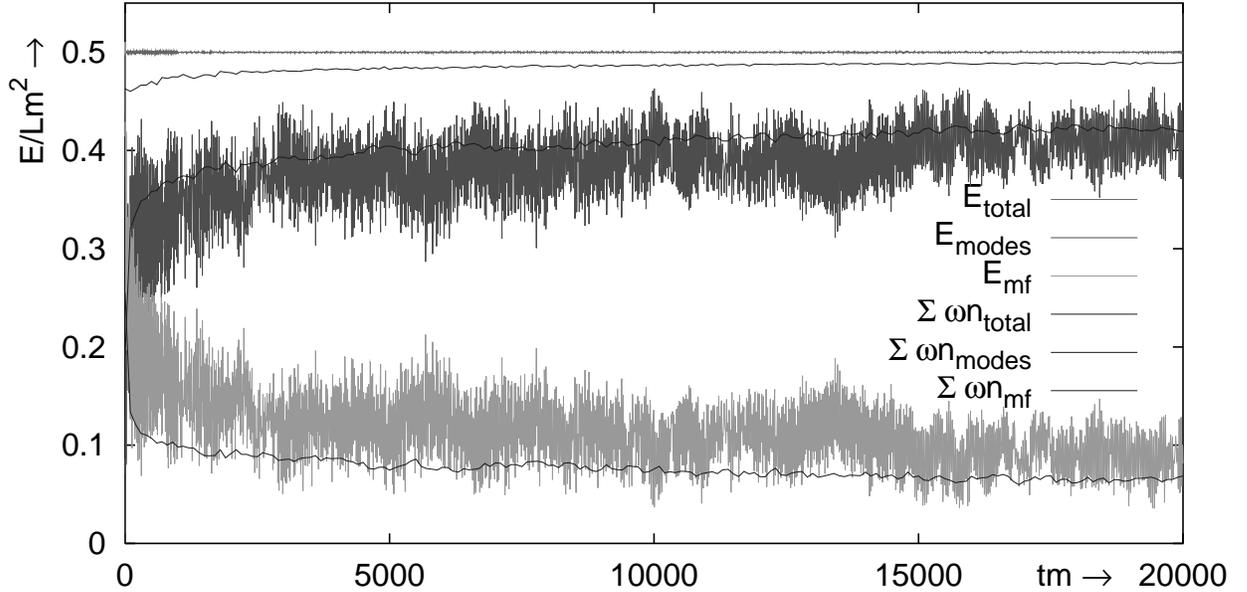}
\caption{The total energy density $E/Lm^2$
(horizontal line at 0.5), energy density of
the mean field (lower band)
and of the modes (higher band).
Also plotted are the various energy densities in the 
quasiparticle interpretation, $\sum_k n_k \om_k/Lm^2$.
\label{fig:mean_late}}
\end{figure}

Initially the mean field carries all the energy in its low momentum modes
$0<k/m\leq \pi/4$ (zero momentum mode excluded). 
Due to interaction with the
inhomogeneous mean field, the modes will not keep the vacuum
form, but get excited. 
Fig.~\ref{fig:mean_late} shows the time dependence of
the energy density for one of the members of the ensemble. 
The total energy is conserved up to a numerical accuracy of about 0.2\%. 
The energy in the mean field  (cf.\ (\ref{Eclas}) for its definition), 
initially equal to the total energy, is decreasing rapidly and
after a time $tm \approx 100$ about 50\% has been transfered to the
modes. The mean field continues losing energy after that time but at a time 
$tm$ of the order 20000 some 15\% is still left.

\begin{figure}[]
\includegraphics[width=\textwidth]{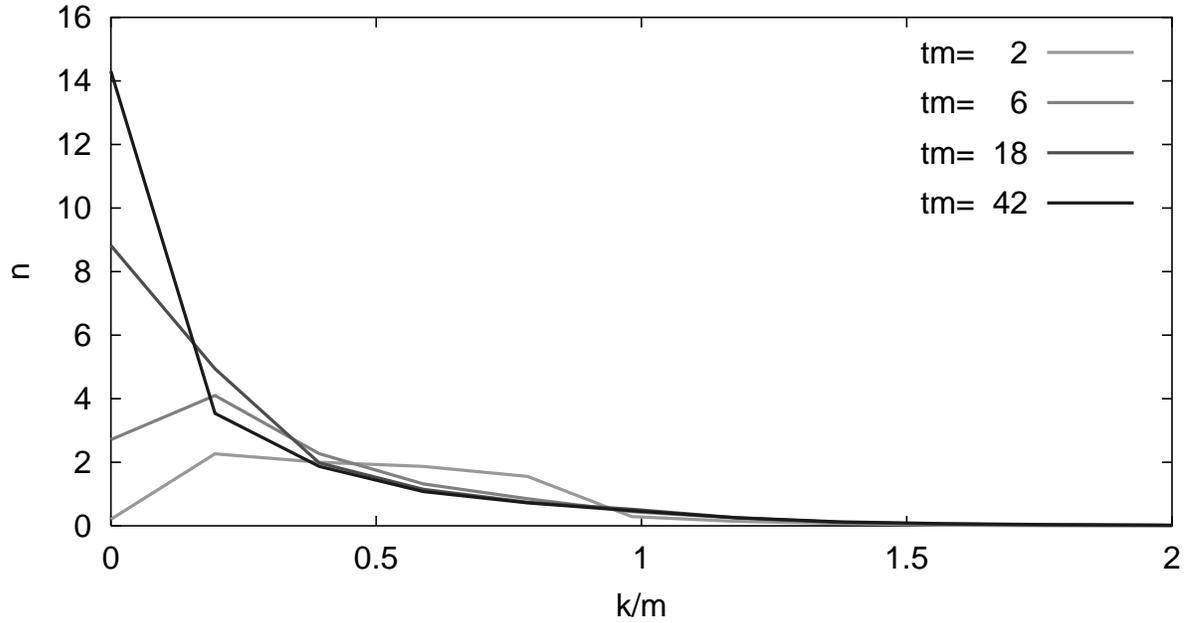}
\caption{Particle number $n_k$ versus $k/m$ for early times. 
\label{fig:nk_early_qc}}
\end{figure}

\begin{figure}[]
\includegraphics[width=\textwidth]{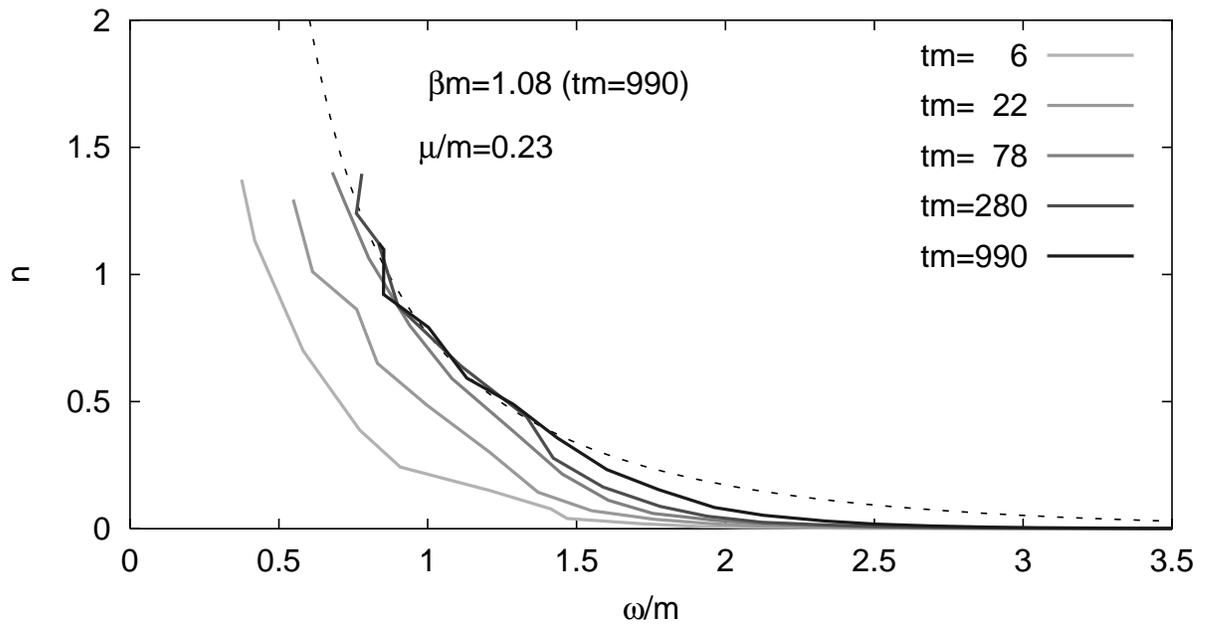}
\caption{Particle number $n_k$
(modes only) versus $\om_k$ for early times. 
\label{fig:n_early_q}}
\end{figure}

The development of the particle numbers $n_k(t)$ at early times is shown
in Fig.~\ref{fig:nk_early_qc},
including the mean field contribution, cf.~(\ref{cq1})--(\ref{cq2}).\footnote{ 
In this and following figures an average is taken over $k>0$ and $k<0$.
The distributions $n_k$ for positive and negative $k$ are equal within 
fluctuations.} 
Initially the mean field gives the main contribution since $n_k^0=0$ for
the modes, but then the mode contribution rapidly takes over.
Because the mean field contribution fluctuates strongly we used as many
as 500 initial conditions for these early times, without coarsening over time.
Fig.~\ref{fig:n_early_q} shows the mode contribution to
$n_k$ as a function of $\om$
(40 initial conditions were used for the data at
$tm > 200$, with no coarsening over time). 
It starts out identically
zero, rises rapidly and then appears to stabilize.
The figure also shows a fit to the Bose-Einstein distribution 
with chemical potential $\mu$ at time $tm = 990$.
A chemical potential is expected to develop temporarily at weak coupling,
since elastic scattering dominates over processes like $2 \leftrightarrow 4$ 
scattering. The fitted temperature ($\bt m = 1.08$) is already approaching
the earlier estimate $T/m \approx 1.1$
based on the energy density.
The complete distribution function 
(including the mean field contribution)
reaches much larger values
at these early times (by a factor 3-4) and the curves appear closer together, 
but the plots are still noisier due to the strongly
fluctuating mean field. 

\begin{figure}[]
\includegraphics[width=\textwidth]{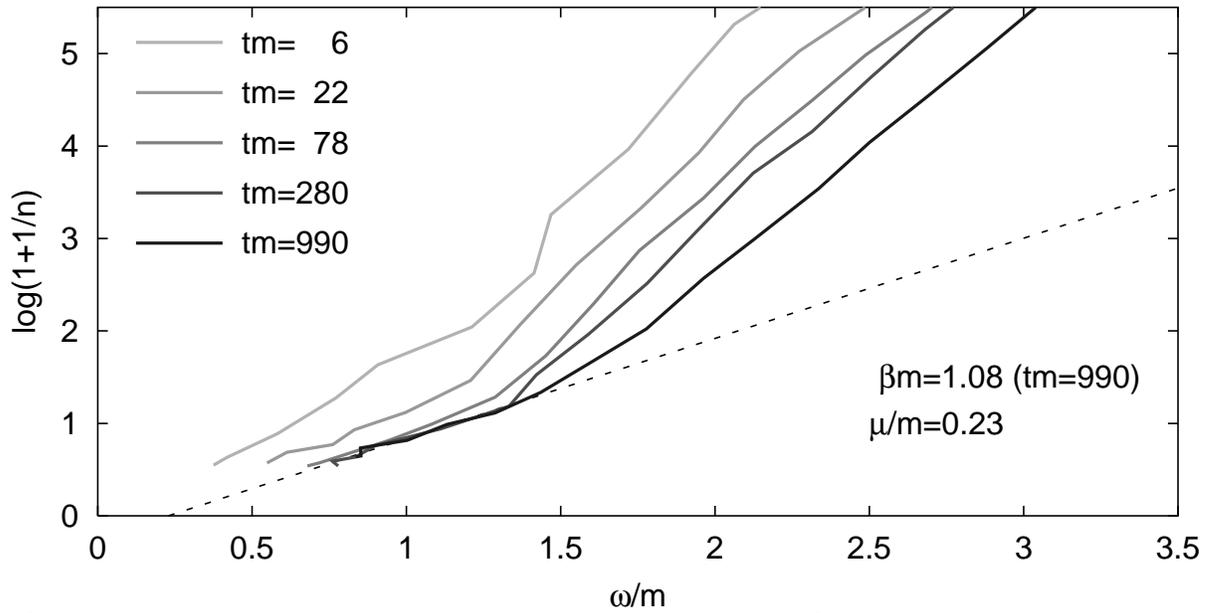}
\caption{Particle number $\log(1+1/n_k)$
(modes only) versus $\om_k$ for early times. The straight line is
a Bose-Einstein fit for the latest time, over $\omega/m < 1.2$.
\label{fig:dist_early_q}}
\end{figure}

\begin{figure}[]
\includegraphics[width=\textwidth]{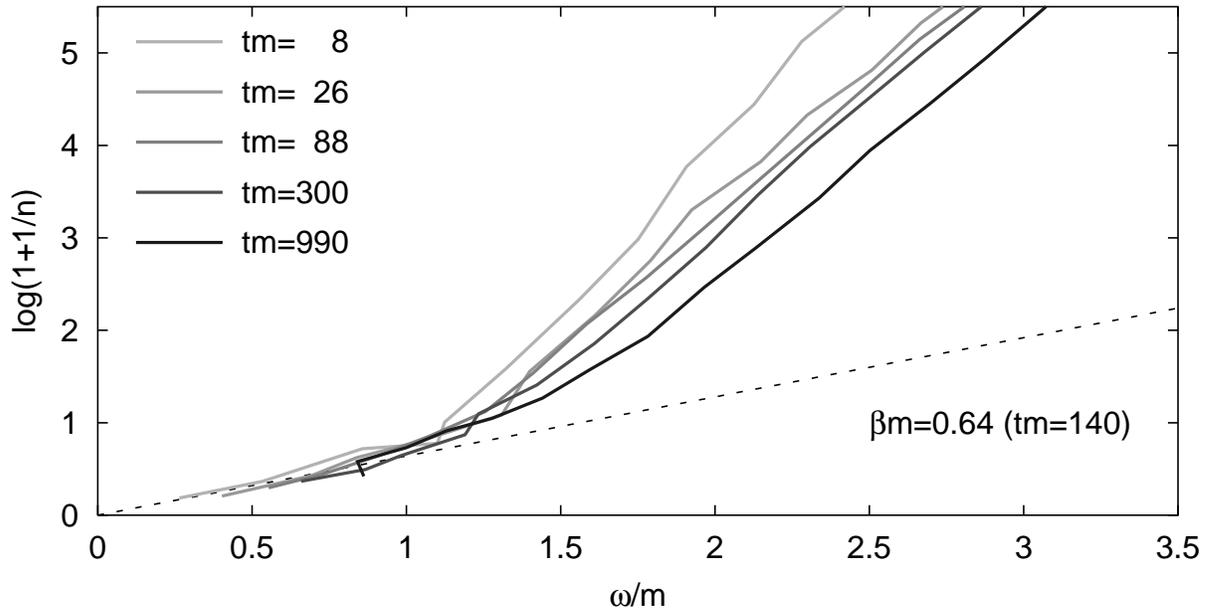}
\caption{As in \protect Fig.~\ref{fig:dist_early_q}, including the mean field
contribution.
\label{fig:dist_early_qc_mu0}}
\end{figure}

\begin{figure}[]
\includegraphics[width=\textwidth]{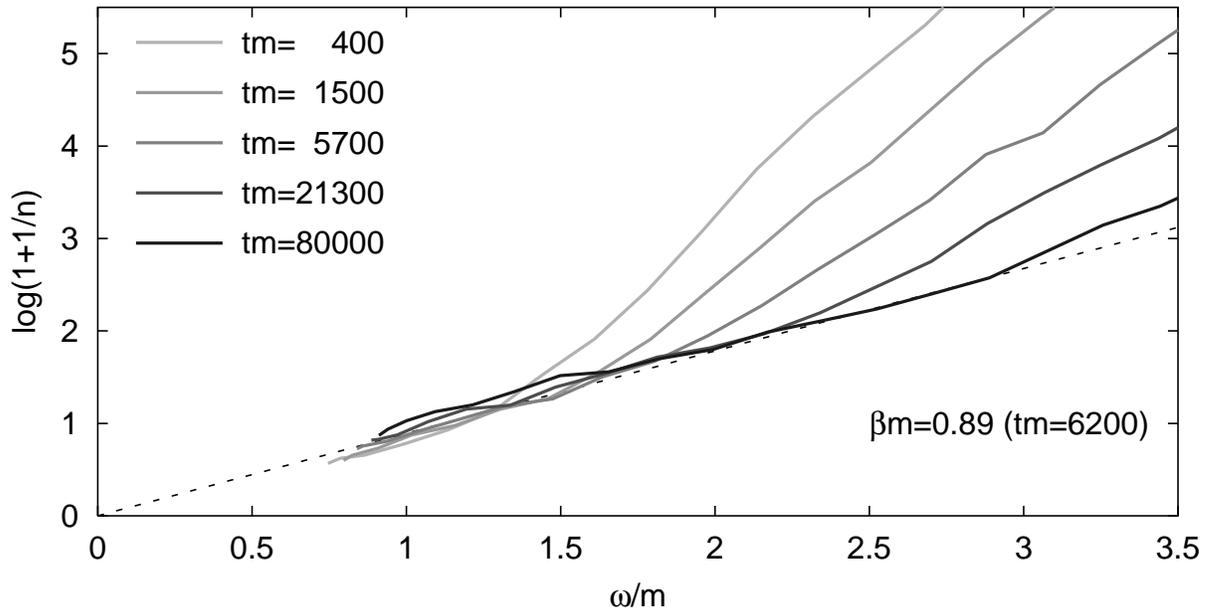}
\caption{The particle numbers (modes only) for later times. 
\label{fig:dist_late_q_mu0}}
\end{figure}

To study the tail of the distribution more easily,
we plot in Fig.~\ref{fig:dist_early_q} $\log(1+1/n)$,
which is linear in $\om$ for a Bose-Einstein distribution. 
We see linear Bose-Einstein behavior developing
at low momenta with gradual participation of the higher momentum modes. 
Including the contribution of the mean field,
shown in Fig.~\ref{fig:dist_early_qc_mu0}, 
we see a more rapid convergence and higher occupation numbers,
giving a higher fitted temperature and smaller chemical potential, 
compared to the data in Fig.~\ref{fig:dist_early_q}.
The trend seen in Figs.~\ref{fig:dist_early_q} and
\ref{fig:dist_early_qc_mu0}
continues at larger times, as shown in Fig.~\ref{fig:dist_late_q_mu0} for
the contribution of the modes only. The plot including the mean field
contribution looks similar. 
(We averaged over a time interval
$tm=24$ (approximately 3.5 oscillation periods) and used only
10 initial configurations.)
The straight line is
a Bose-Einstein fit with zero chemical potential at $tm = 6200$
in the region $\omega/m < 1.8$.
We see that the slope
is roughly constant in time and that the
thermalized part of the distribution is extending to higher values of $\omega$,
roughly linear in $\log tm$. 

\begin{figure}[]
\includegraphics[width=\textwidth]{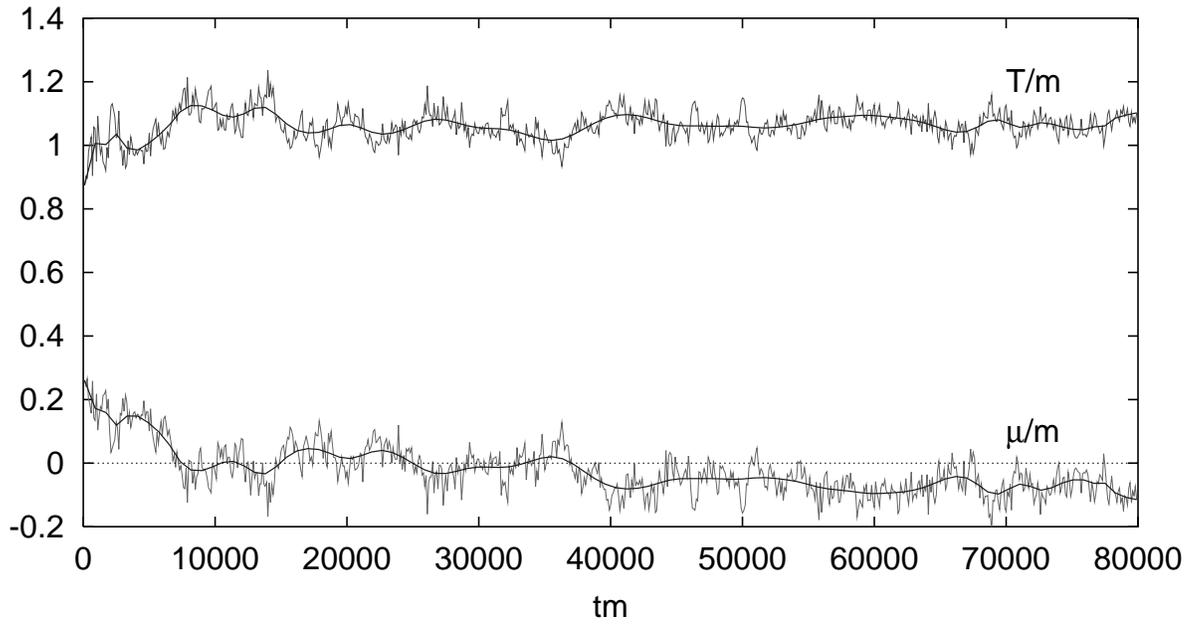}
\caption{The Bose-Einstein temperature for particle numbers plotted in
Fig.~\ref{fig:dist_late_q_mu0}. 
The smoother lines are drawn to guide the eye.
\label{fig:beta_late}}
\end{figure}

\begin{figure}[]
\includegraphics[width = \textwidth]{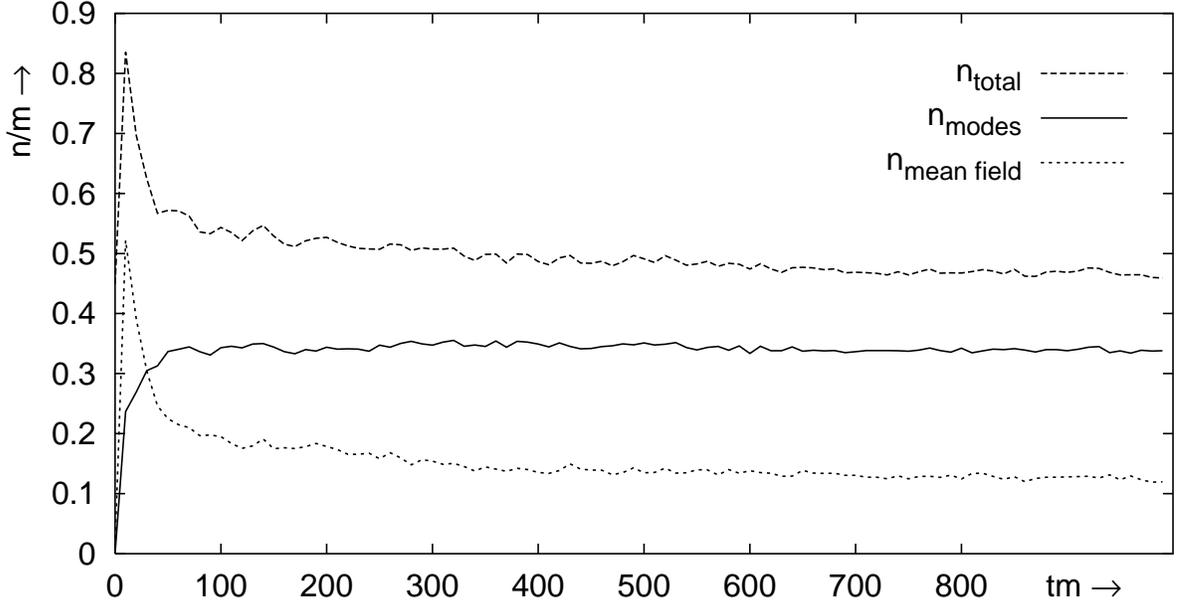}
\caption{Particle densities $n/m = \sum_k n_k/Lm$.
\label{fig:total_n}}
\end{figure}

In Fig.~\ref{fig:beta_late} a plot is made of the 
Bose-Einstein temperatures from the fits (modes only) as a
function of time. 
For times $tm < 3000$ the fit is
made over the interval $\omega/m < 1.4$ while for later times this
is increased to $\omega/m < 1.8$. 
The figure shows an anticorrelation between $T$ and $\mu$ which
would be meaningful (i.e.\ not just a fitting artifact)
if the particle density $n=\sum_k n_k/L$ is constant (or has evidently smaller
fluctuations).
This seems to be the case indeed: as shown in Fig.~\ref{fig:total_n}, the
density $n$ corresponding to the modes only is quite constant for times 
$tm > 100$, and in fact continues to remain so up to times of over $5000$. 
On a larger time scale  of order 10000 or so it drops somewhat.
The initial 
approach of $n/m$ (modes only) to the value $\approx 0.34$ can be fitted
to an exponential, which yields an equilibration time scale $\tau m = 15$ -- 20,
depending on the fitting range.

We have to be careful, however, that our $\mu$ is not an artifact
of the fitting procedure.
We believe this to be the case for the larger times $tm \agt 40000$
where $\mu$ goes negative.
As can be seen (with difficulty) in Fig.~\ref{fig:dist_late_q_mu0}, 
the distribution starts to deviate at low $\om$ upwards
from the straight line, corresponding to a suppression of $n_k$ compared to
the Bose-Einstein form. We interpret this as a contamination by classical
behavior $n_k \approx T_{\rm cl}/\om_k$, as in (\ref{nclas}), as will be
argued later in this section.

Let us now compare with analytical results derived from
the equilibrium finite temperature effective potential (\ref{eqeffpotT}).
Around $tm=15000 \dots 20000$ the temperature measured in the simulation
is $T/m = 1.1$. The effective potential then gives for the thermal mass
$m(T=1.1)/m=v(T=1.1)/v=0.93$. 
We derive the thermal mass in the simulation from the dispersion
relation of measured $\omega_k$.
It is in very good agreement with a free form: 
$\om_k^2 = m^2(T) + k^2$. 
A straight line fit of $\om^2$ versus $k^2$
over the interval $tm=15000 \dots 20000$ gives a slope $1.00$
and an offset 
$m(T=1.1)/m=0.908$. This is also in good agreement with the volume
average of the mean field, which is $0.91$. 
(These values are somewhat lower than the position of the minimum in
the effective potential because of its asymmetric shape, but the difference is
small because of the small amplitude of the mean field oscillations.) 

The quasiparticle aspect can be investigated further by looking at the 
energy $\sum_k n_k\om_k$, as plotted in Fig.~\ref{fig:mean_late}. We have made a
distinction between the particle number as derived from the mean field,
quantum and total two-point function. We see that the total 
energy in the particles (mean field + modes)
is only a few percent lower than the 
total energy is the system, 
as may be expected for a weakly coupled system. 
%
It is also interesting to note that the quantum modes thermalize with
the same temperature 
1.1 $m$ the system would have if all energy 
would be distributed according to a Bose-Einstein distribution with zero
chemical potential, although the modes
carry initially much less than the total energy.

We now turn to the very long time behavior of the system, where we expect
Bose-Einstein behavior to be replaced by classical
equipartition according to the effective hamiltonian~(\ref{heff}). The
numerical computation of 
the equilibrium distribution functions in this regime is very
difficult as it changes exceedingly slowly
(cf.\ the slow $\log t$-like
population of the high momentum modes in Fig.~\ref{fig:dist_late_q_mu0}).
We therefore have carried out simulations in
a smaller system at stronger coupling and at larger energy densities
in order to make time scales a lot shorter.
Here we present data for $N=16$, $Lm=1$, $\lambda/m^2=1$ and $E/Lm^2=36$,
for which the system is in the `symmetric phase'. 
In Fig.~\ref{fig:dist_extreme} we plotted $n_k\omega_k$ 
(modes + mean field) versus the integer $kL/2\pi = k/2\pi m$, 
for different times. Note that we needed to excite initially
also the highest momentum modes, 
otherwise the system would not reach final equilibrium sufficiently closely
even after a time of 12 million. 
Classical equipartition suggests $n_k \omega_k =T_{\rm cl}$, 
giving a straight horizontal line in the plot. 
We see indeed flat behavior, with lower momentum modes tending to 
have somewhat smaller occupation numbers, except for the zero mode. 
Runs at small coupling 
$\lambda/m^2=1/12$ in larger volumes $Lm=4$ and $Lm=16$ 
in the `broken phase' showed similar results,
except that the zero modes were less exceptional.

So we do find approximate classical $n_k = T_{\rm cl}/\om_k$ behavior at 
very large times.
Classical equipartition
leads to small temperatures $T_{\rm cl} = O(1/N)$. If this behavior sets
in first for the low momentum modes, then these will appear to be under-occupied
compared to the Bose-Einstein distribution at temperature $T > T_{\rm cl}$.
This is indeed the trend noticed earlier in Fig.~\ref{fig:dist_late_q_mu0}, 
where the low momentum data at times $tm > 20000$
lie above the straight line going through the data at larger momenta.


\section{Damping rate}
\label{secdamp}
In the previous section we have seen the system equilibrate initially 
on a time scale of $tm = 15$ -- 20 in its low momentum modes, 
with a particle distribution approaching the Bose-Einstein form.
Subsequently this approach progressed rather more slowly towards 
higher momenta, on a time scale which is hard to quantify, of the
order of thousands to tens of thousands. 
To get more information in this regime we turn to autocorrelation functions. 
For a homogeneous ensemble at finite temperature,
the spatial Fourier transform $F_k(t)$ of   
the symmetrized autocorrelation function 
\be
F_k(t-t') = \int dx\, e^{-ik(x-x')}\,
\left[\half\langle \{\hat\varphi(x,t),\hat\varphi(x',t')\}\rangle
-\langle\hat\varphi(x,t)\rangle\langle\hat\varphi(x',t')\rangle\right]
\ee
is given in terms of the spectral function 
by standard formulas.
In case of weak coupling the spectral function is expected to exhibit a
strong peak around the  
mass shell of the quasiparticles, which leads to exponential decay
of $F_k(t)$ in an intermediate time regime. The decay rate is
called `the plasmon damping rate'.

In the Hartree ensemble 
approximation $F_k(t)$ can be written as the sum of a mean field
part and a contribution from the mode functions. 
It is easiest to compute the mean field part. This would give no
information in case of constant mean fields, since it would be identically zero.
However, 
we expect mean field and modes to be sufficiently coupled
to gain useful information on the damping rate from the mean field part only.
Even at late times $tm = 30000 -80000$ we observed the
back reaction $3\lm \sum_{\al} |f_{\al}(x,t)|^2$ of the modes on the mean field
to be strongly fluctuating in space and time. Fluctuations in the
modes will then cause corresponding fluctuations in the mean field.

\begin{figure}[]
\includegraphics[width=\textwidth]{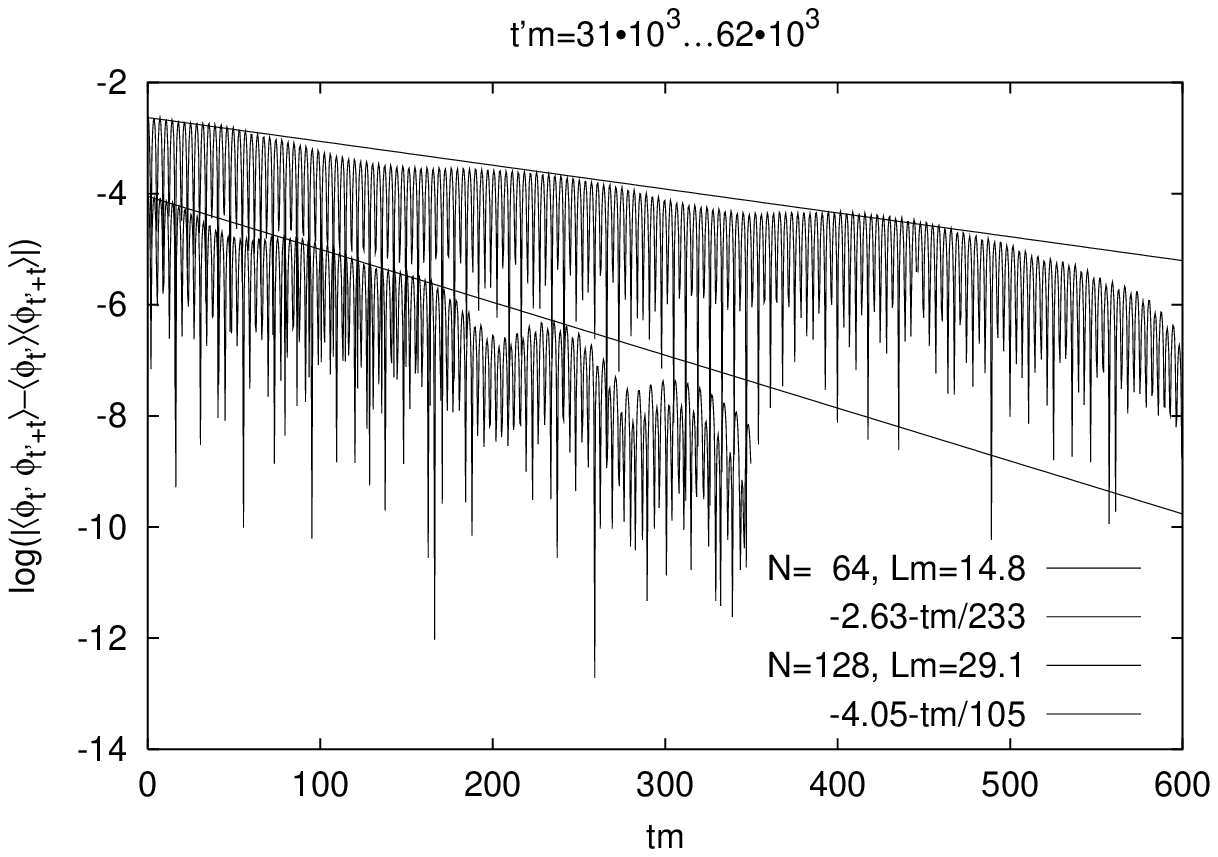}
\caption{Numerically computed auto-correlation functions $\log|F_{0\rm mf}(t)|$
versus time $tm_T$, with $m_T$ the temperature dependent mass.
The coupling is weak, $\lm/m_T^2 = 0.11$ and the temperature $T/m_T\approx 1.4$
for the smaller volume (with significant deviations from the Bose-Einstein
distribution) and $\approx 1.6$ for the larger volume (reasonably BE).
}
\label{fdamp}
\end{figure}

We have computed the mean field part $F_{0\rm mf}(t)$ at $k=0$,
obtained by taking a time average 
after an initial equilibration period $t \in(0,t_0)$:
\bea
F_{0\rm mf}(t) &=&
\frac{1}{(t_1-t_0)}\int_{t_0}^{t_1}dt'\,\tilde\varphi_0(t+t')
\tilde\varphi_0(t')
\nonumber\\&&\mbox{}
-\frac{1}{(t_1-t_0)^2}\int_{t_0}^{t_1}dt'\,\tilde\varphi_0(t+t')
\int_{t_0}^{t_1}dt'\,\tilde\varphi_0(t').
\eea
where $\tilde\varphi_0(t) = \int dx\, \varphi(x,t)/\sqrt{L}$.
No average was taken over initial conditions.
Fig.~\ref{fdamp} shows two examples of $F_{0\rm mf}(t)$, for which the average
was taken after an equilibration time of $t_0 m \approx 31000$ over the
interval $(t_0 m,t_1 m) \approx (31000,62000)$. 
We see roughly exponential decay modulated by oscillations.
At first the oscillations looked suspicious to us, 
as if there were strong memory effects and no damping, but other simulations
{\em with} averaging over initial conditions (this time in the symmetric phase)
gave similar results. As a check we
used two-loop perturbation theory to
calculate the spectral function  
in the full (not Hartree approximated) theory. To our surprise this led to
similar oscillations, modulating exponential decay.
The reason is that in one space dimension collinear divergences 
lead to a spectral function with {\em two adjacent peaks}
\cite{SaSmVi00ab}. 
So we conclude that the damping behavior
in Fig.~\ref{fdamp} is real. The straight lines
indicate damping times $\tau m_T \approx 105$ and $\approx 233$. We
use the finite temperature mass here to set the scale as this appears naturally
in resummed perturbation theory.
For the first example (with the larger volume) the 
corresponding particle distribution was found to be reasonably of
the Bose-Einstein form, with zero chemical potential and 
temperature $T/m_T \approx 1.6$. The two loop perturbative calculation 
gives a $\tau m_T \approx 67$ for this temperature, 
which we consider encouragingly
close to the Hartree ensemble result $\approx 105$. 
We should however warn the reader that
the numerical computation of autocorrelation functions is quite difficult and
that there may be large {\em statistical} errors in the numbers given.

\section{Discussion}
\label{secconcl}
We presented results of simulations mainly for a weakly coupled system,
such that near equilibrium a description in terms of quasiparticles 
is expected to be reasonable (we will check this expectation in a future
publication \cite{SaSmVi00c}).
Starting with distributions which are initially far out of equilibrium, in  
which only low momentum modes $k\lesssim m$ of the classical field were excited
with low energy density,
we observed approximate thermalization with a particle distribution function
approaching the Bose-Einstein form.
After a fairly rapid initial thermalization at low momenta, 
the gradual adjustment of progressively higher momentum modes
is very slow.
The energy in the mean field gets transfered to the two-point function and
one might think that the system behaves as if the mean field were
constant. However, this is not the case: up to large times $tm = 80000$
the mean field keeps fluctuating in space and time and carries a non-negligible
fraction of the total energy.
Correspondingly, there is a `plasmon damping rate', which turns out to be 
similar in magnitude to that
predicted by two-loop perturbation theory (with no further gaussian
approximation).
It is hard to assign a time scale for the gradual adjustment of the
distribution at higher momenta, but it
appears to be at least two orders of magnitude larger than the 
equilibration time $\tau m \approx 20$ for the particle density, found 
at early times ($tm = O(10)$), or for the 
damping time $\tau m \approx 100$ for the zero mode of the mean field,
found at larger times ($tm = O(10000)$.
Slow thermalization was also found in a recent study of the fully nonlinear
classical system in the symmetric phase \cite{AaBoWe00b}. 
Using our parameter combination
$\lm T/m^3 \approx 1.1/12$ in their empirical fit 
$1/\tau m = 5.8\, 10^{-6}\,(6 \lm T/m^3)^{1.39}$ would give 
$\tau m \approx 4\, 10^5$.

On a large time scale, perhaps of the order of $tm = 10000$ or more
the distribution moves away from the quantum (Bose-Einstein) form
towards classical equipartition. We never reached this classical
equipartition for the weak coupling and low temperature used in this study.
It would have taken much too long.
Only for very small systems at high energy density and/or coupling
we were able to reach a situation resembling
classical equipartition.

We have carried out many more simulations at higher energy densities,
and larger couplings, in which the approximate
quantum nature of the distribution at intermediate times was also evident. 
With higher energy density and/or larger coupling 
the effective coupling strength $n_k\lm/m^2$ increases.
Things then go quicker and the time scales of quantum
versus classical equilibration get closer and might even get blurred. 
Furthermore, the Bose-Einstein distribution,
on which we based our analysis, might get distorted 
by nonperturbative effects.
We may have seen such effects already 
in a significant enhancement of $n_k$ at low momenta, 
in simulations at larger volume. 

We have also performed simulations in the `symmetric phase' of the model.
The picture there is confusing. At similar couplings and initial conditions
as described in the previous sections nothing much seems to happen. 
Presumably, the reason is the extremely short range nature of
the pure $\varphi^4$ interaction in the `symmetric phase'.
In the `broken phase' there is
also a non-zero three-point coupling, giving the interactions
a finite range.  But at higher energy 
density and/or coupling there seems to be hardly a time regime in
which the distribution function looks sufficiently Bose-Einstein.

Summarizing, on the one hand our intuitive expectation that there may be quantal
thermalization in the gaussian approximation, due to scattering
of the mode particles via the arbitrary inhomogeneous mean field,
appears to be validated, but on the other hand it is not 
clear how useful this approximation can be for equilibrium physics,
e.g.\ at finite density. It is possible that starting closer to
quantum thermal equilibrium the time to reach thermalization is reduced and
the intermediate time regime of quantal equilibrium can be stretched to
do useful computations. Then it will be interesting to compare 
the gaussian approximation with the classical approximation and see
which one fares best. We will address this aspect in a separate
paper \cite{SaSmVi00c}, where we will also investigate the possibility of
using fewer mode functions.\footnote{The numerical cost of the 
inhomogeneous gaussian approximation is substantial 
and scales like $N^{2d+1}$ for an $N^d$ spatial lattice.}
\\

Acknowledgements\\
We thank Gert Aarts, Bert-Jan Nauta and Chris van Weert 
for useful discussions. This work is supported by FOM/NWO.

\appendix
\section{The diagonal coherent state representation}
To derive the representation (\ref{rhtocoh}) consider
first a quantum mechanical system of two degrees of freedom with
canonical variables $p$ and $q$. 
Let $|pq\rangle$ be a normalized coherent state, such that
\bea
\ah|pq\rangle &=& \frac{1}{\sqrt{2\om}}\, (\om q + i p)\, |pq\rangle,
\;\;\;\;
\ah \equiv \frac{1}{\sqrt{2\om}}\, (\om \qhat + i \phat),
\nonumber\\
\langle p'q'|pq\rangle &=& \exp\left\{
\frac{i}{2}\,(pq'-p' q) 
-\frac{1}{4\om}[ \om^2 (q-q')^2 + (p-p')^2] 
\right\}
\nonumber\\
\int \frac{dp\,dq}{2\pi}\, |pq\rangle\langle pq| &=& \hat{1}.
\eea
where $\om>0$ is arbitrary. As is well known, the coherent states form a
(over-complete) set, so it should be possible to represent an
arbitrary operator $\hat\rh$ in the form
\be
\hat{\rh} = \int \frac{dp\,dq}{2\pi}\, \rh(p,q)\,|pq\rangle\langle pq|.
\ee
In our application $\hat\rh$ is a density operator, for which
\be
\int \frac{dp\,dq}{2\pi}\, \rh(p,q) = 1.
\ee
Taking
matrix elements of the above equation with
$|p',q'\rangle$ and $\langle -p',-q'|$ gives
\be
e^{(\om^2q^{\prime 2} + p^{\prime 2})/2\om}\,
\langle -p',-q'|\hat\rh|p',q'\rangle
= \int \frac{dp\,dq}{2\pi}\,
e^{i(p' q - pq')}\, 
e^{-(\om^2 q^2 + p^2)/2\om}\,
\rh(p,q), 
\ee
from which follows that
the function $\rh(p,q)$ is given by the inverse Fourier transform
\be
\rh(p,q) = e^{(\om^2 q^2 + p^2)/2\om}\int \frac{dp'\, dq'}{2\pi}\,
e^{-i(p' q - pq')}\, 
e^{(\om^2q^{\prime 2} + p^{\prime 2})/2\om}\,
\langle -p',-q'|\hat\rh|p',q'\rangle.
\ee

A trivial example is a coherent state centered about $p_1$, $q_1$, for which
$\rh(p,q) = 2\pi\dl(p-p_1)\dl(q-q_1)$.  Another simple 
example is given by the thermal density operator of the harmonic
oscillator with hamiltonian $H = (\om^2 q^2 + p^2)/2$,
\be
\hat\rh = \frac{1}{Z}\,\exp\left[-\bt\om\left(\ahd\ah + \half\right)\right],
\ee
with $Z$ the partition function, such that $\Tr\hat\rh = 1$.
Choosing the $\om$ in the definition of the coherent
states equal to the $\om$ appearing in this $\hat\rh$,
it follows that
\be
\langle -p',-q'|\hat\rh|p',q'\rangle = 
\frac{1}{Z}\,\exp\left[-\left(e^{-\bt\om}+1\right)
\frac{1}{2\om} \left(\om^2 q^{\prime 2} + p^{\prime 2}\right) 
- \half\bt\om\right],
\ee
and
\be
\rh(p,q) = \frac{1}{Z} \exp\left[-\left(e^{\bt\om}-1\right) 
\frac{1}{2\om} \left(\om^2 q^{2} + p^{2}\right) 
+ \half\bt\om\right].
\ee
We recognize the inverse Bose-Einstein distribution, 
$\exp(\bt\om)-1$, in the exponent. 
For large temperatures, $\bt\om \ll 1$,
$\rh(p,q)$ approaches the classical Boltzmann distribution 
$\exp(-\bt H)$. 
In the limit of zero temperature we get
the distribution representing the ground state,
\be
\rh(p,q) = 2\pi\dl(p)\dl(q).
\ee
More examples can be found in \cite{MaWo91}.
The generalization to the scalar field is straightforward.

\section{Equipartition?}

The effective hamiltonian $H_{\rm eff}[\varphi,\pi,\xi,\et]$ 
of the gaussian approximation
is conserved in time. So one may expect that after very large times
the system reaches {\em classical} equilibrium.
Assuming ergodicity, time averages will then correspond to
the Boltzmann distribution $\exp(-H_{\rm eff}/T)$, 
under the constraints of the conserved
generalized angular momenta $L_{\al a,\bt b}$ (cf.\ (\ref{lang})).
We shall now derive an approximate form for the
particle distribution function $n_k$ corresponding to this classical
equilibration. 

In our derivation we assume the system to be weakly coupled, 
such that
we may approximate $H_{\rm eff}$ in the Boltzmann distribution by a
free field form (possibly after having shifted $\varphi$ by its
equilibrium value $v$ such that $\langle\varphi\rangle = 0$),
\be
H_{\rm free} = \int dx\,\left[
\half\pi^2 + \half (\partial\varphi)^2 + \half m^2 \varphi^2  +
\sum_{\al} (|\et_{\al}|^2 + |\partial\xi_{\al}|^2 + m^2 |\xi_{\al}|^2)
\right],
\ee
where $m$ is an effective mass. For convenience we use a complex
formalism for the mode functions ($\xi_{\al} = 
(\xi_{\al 1} -i \xi_{\al 2})/\sqrt{2}
=\sqrt{n^0_{\al} + 1/2}\, f_{\al}$, 
cf.\ (\ref{xiintro})).\footnote{We added a superscript 0 to
$n_{\al}$ to indicate that these are the initial values
at time $t=0$, in order to avoid
possible confusion with the $n_k$.}
The generalized angular momenta are just the naturally
conserved charges of the complex fields,
\be
Q_{\al} = i\int dx\, (\xi^*_{\al}
\et^*_{\al} - \et_{\al} \xi_{\al})= L_{\al 1,\al 2} = n^0_{\al} + \half.
\ee
We take them into account by introducing chemical potentials
$\mu_{\al}$, such that the average charges are equal to their
values set by the initial conditions, $Q_{\al} = n^0_{\al}+1/2$.
It is not immediately clear that this procedure is correct, because 
these initial values are not extensive and therefore relative fluctuations
will be large, but the emerging formulas below look reasonable. 
Imposing the constraints exactly appears to be quite cumbersome, except
for $N=1$. Recall that $N$ is the number of complex mode functions, which
in the lattice regularization is equal to the number of lattice sites:
$N=\sum_k = \sum_{\al}$.
Here we shall assume a sharp momentum cutoff $|k| < \Lm$, for simplicity.
 
The classical grand canonical average will be indicated by an over-bar:
\be
\overline{F} = \frac{1}{Z_c}
\int [d\varphi\,d\pi][\prod_{\al}d\xi_{\al}\,d\et_{\al}]\,
\exp\left[-\frac{1}{T}
\left(H_{\rm free}-\sum_{\al} \mu_{\al} Q_{\al}\right)\right]\, F,
\ee
with $Z_c$ the partition function such
that $\overline{1} = 1$. 
Our approximation for $n_k$ is now given by
($\om_k = \sqrt{m^2 + k^2}$)
\bea
S(x,y) &=& \frac{1}{L} \sum_k e^{ik(x-y)}\,
\frac{n_k + 1/2}{\om_k},
\nonumber\\&=&
\overline{\varphi(x)\varphi(y)} + \sum_{\al}\left[
\frac{n^0_{\al}+1}{n^0_{\al} + 1/2}\,
\overline{\xi_{\al}(x)\xi_{\al}^*(y)} +
\frac{n^0_{\al}}{n^0_{\al} + 1/2}\,
\overline{\xi_{\al}^*(x) \xi_{\al}(y)}\right].
\eea
The calculation is a straightforward free field exercise.
Introducing the classical analogues of the creation and
annihilation operators,
\be
\varphi(x) = \sum_k \frac{e^{ikx}}{\sqrt{2\om_k L}}\,(a_k + a^*_{-k}),
\;\;\;\;   
\xi_{\al} =\sum_k \frac{e^{ikx}}{\sqrt{2\om_k L}}\,(a_{\al k} +
b^*_{\al -k}),
\ee
and accordingly for the canonical momenta $\pi$ and $\et_{\al}$,
we get
\bea
H_{\rm free} &=& \sum_k \left[|a_k|^2 + \sum_{\al}\left(|a_{\al k}|^2 +
|b_{\al k}|^2\right)\right] \, \om_k,
\nonumber\\
Q_{\al}&=&\sum_k \left[|a_{\al k}|^2 - |b_{\al k}|^2\right].
\eea
It follows that
\bea
n_k +\frac{1}{2} &=& \overline{|a_k|^2} + \sum_{\al} \left(\
\overline{|a_{\al k}|^2} + \overline{|b_{\al k}|^2}\right)
\nonumber\\&=&
\frac{T}{\om_k} + \sum_{\al}\left(\frac{T}{\om_k-\mu_{\al}} +
\frac{T}{\om_k + \mu_{\al}}\right).
\label{nmual}
\eea
The $\mu_{\al}$ are to be determined by the conditions 
\bea
n^0_{\al} + \half &=& \overline{Q_{\al}} = \sum_{k}\left(
\overline{|a_{\al k}|^2} - \overline{|b_{\al k}|^2}\right)
\nonumber\\&=&
\sum_k \left(\frac{T}{\om_k - \mu_{\al}} - \frac{T}{\om_k + \mu_{\al}}\right).
\label{mualeq}
\eea

Before turning to the case $n^0_{\al} = 0$ used mostly in this paper, we
comment on the properties of the above equations. 
Suppose there is only one complex mode
function (`quantum mechanics'): $N=1$. Then
the solution of the equations is given by 
\bea
\mu &=& \sqrt{\om^2 + \frac{T^2}{(n^0 + 1/2)^2}} - \frac{T}{n^0 + 1/2},
\nonumber\\
n+\half &=& \sqrt{\left(n^0 + \half\right)^2 + \frac{T^2}{\om^2}} + \frac{T}{\om},
\eea
for which $n \geq n^0$. We see that $\mu \to \om$, $n\to n^0$ 
as $T\to 0$, and $\mu\to 0$, $n\to\infty$ as $T\to\infty$.

For finite $N$ 
Eq.\ (\ref{mualeq}) for $\mu_{\al}$ can be rewritten as a polynomial equation
of degree $2N$ by multiplying the 
LHS and RHS by $\prod_k (\om_k^2 - \mu_{\al}^2)$.
So there are in principle $2N$ solutions for each $\mu_{\al}$. For $T\to 0$ 
we have a solution in which $\al \leftrightarrow k$ (as in (\ref{modevac}),
behaving as  
\be
\mu_k = \om_k - T/(n_k^0 + 1/2) + \cdots, 
\;\;\;\;
n_k = n^0_k + \cdots.
\label{mutom}
\ee

For the case $n^0_{\al} \equiv 0$ it is natural to look for a solution in 
which all the chemical potentials are equal, $\mu_{\al}=\mu$.
Eq.\ (\ref{mualeq}) then reduces to
\bea
\half &=& 2T\mu\sum_k \frac{1}{\om_k^2 - \mu^2}
\approx 2TL \mu\int_0^{\Lm} \frac{dk}{\pi}\, \frac{1}{m^2 + k^2 - \mu^2}
\nonumber\\
&\approx&
\frac{TL\mu}{\sqrt{m^2 - \mu^2}},
\label{Qtomu}
\eea
for large volumes $mL\gg 1$ and large momentum cutoff $\Lm/m \gg 1$ 
(the integral converges for $\Lm\to\infty$.) It follows that 
\be
\mu \approx \frac{m}{\sqrt{1 + 4 T^2 L^2}}.
\ee
On the other hand, we have from (\ref{nmual}),
\be
n_k + \half = \frac{T}{\om_k} + \frac{2 NT\om_k}{\om_k^2 - \mu^2},
\label{nmu}
\ee
which depends explicitly on the number of modes $N$. We see that $n_k+ 1/2$ 
falls roughly like $1/\om_k$, and there is a danger that $n_k$ may get
negative for large $\om_k$, which should not happen. 

\begin{figure}[]
\includegraphics[width=\textwidth]{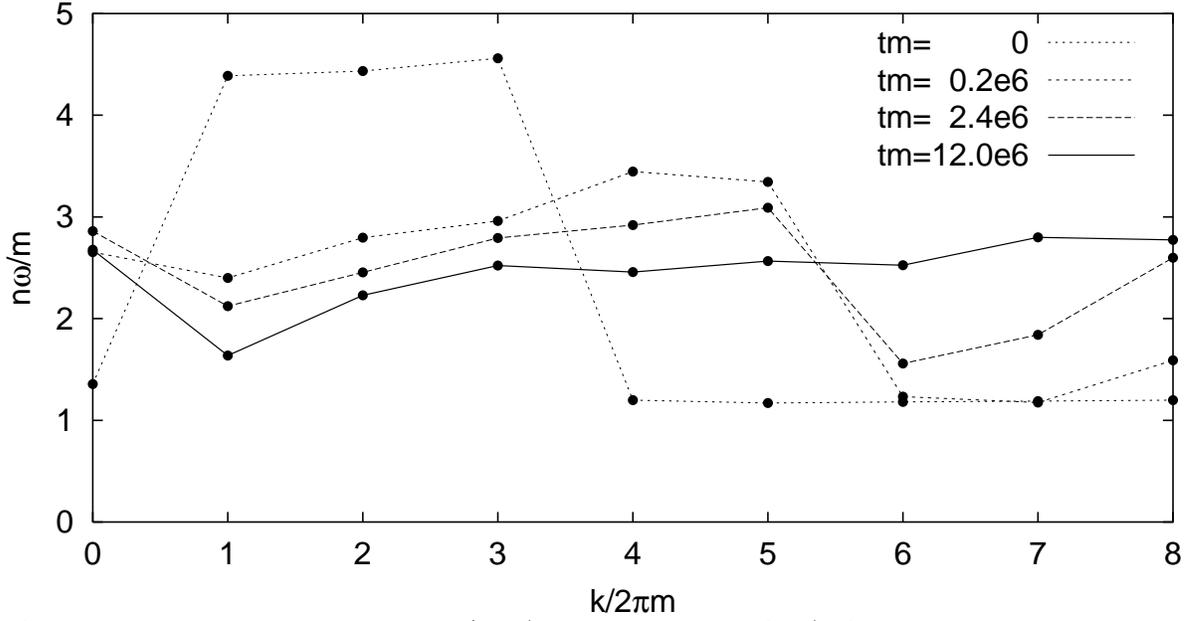}
\caption{Energy distribution $n_k\om_k/m$ (modes + mean field) 
for a small system with $N=16, Lm=1, E/Lm^2=36$. 
\label{fig:dist_extreme}}
\end{figure}

In fact, in our numerical simulations
we always found the $n_k$ to be positive, but
not following the distribution (\ref{nmu}) for all $k$. Even after
very large times we usually found that only a limited
number of modes are able to thermalize approximately classically, except
for small systems such as in Fig.~\ref{fig:dist_extreme}.
%

If we approximate $N = \sum_k\approx L\int_0^{\Lm} dk/\pi = L\Lm/\pi$,
$\om_{\Lm} \approx \Lm$, the condition
$n_{\Lm} + 1/2 \approx 2TN/\Lm \geq 1/2$ leads to $LT \geq \pi/4$, 
If this condition is not
satisfied, more complicated solutions for the chemical potentials may be needed
in which $\mu_k\approx\om_k$, as in (\ref{mutom}). 
We have explored such solutions on the lattice, using Mathematica. 
Despite ambiguities (e.g.\ funny behavior of the alternating lattice modes), 
such solutions indicate that $n_k \om_k$ is quite constant
(but apparently not exactly), i.e.\ approximate equipartition.

So we tentatively conclude that, approximately, $n_k \approx T_{\rm cl}/\om_k$
is the predicted form for the particle distribution at very large times.

%
%

%
%
%

\begin{references}
\bibitem{BoMoRu00} 
                 D.~B\"odeker, G.D.~Moore, K.~Rummukainen,
                 Phys.~Rev.~D61 (2000) 056003, hep-ph/9907545.
\bibitem{SmTa99}
                W.H.~Tang, J.~Smit,
                Nucl.~Phys.~B540 (1999) 437, hep-lat/9805001.
\bibitem{afterpreh} 
           S.~Yu.~Khlebnikov, I.I.~Tkachev, Phys.~Rev.~Lett.~77 (1996) 219;
           T.~Prokopec, T.~Roos, Phys.~Rev.~D55 (1997) 3768, hep-ph/9610400;
           G.~Felder, L.~Kofman, Phys.~Rev.~63 (2001) 103503, hep-ph/0011160.
\bibitem{newewbary}
           J.~Garc\'\i a-Bellido, D.Y.~Grigoriev, A.~Kusenko, M.~Shaposhnikov,
           Phys.~Rev.~D60 (1999) 123504, hep-ph/9902449.
           A.~Rajantie, P.M.~Saffin, E.J.~Copeland, 
           Nucl.~Phys.~B587 (200) 403, hep-ph/0012097.

\bibitem{Pa97} G.~Parisi, Europhys. Lett 40 (1997) 357; 
\bibitem{AaBoWe00a}
             G.~Aarts, G.F.~Bonini, Ch.~Wetterich, 
             Nucl.~Phys.~B587 (2000) 403, hep-ph/0003262.
\bibitem{AaBoWe00b}
               G.~Aarts, G.F.~Bonini and Ch.~Wetterich, 
               Phys.~Rev.~D63 (2001) 025012, hep-ph/0007357. 
\bibitem{AaSm34} 
            G.~Aarts, J.~Smit,
            Phys.~Rev.~D61 (2000) 025002, hep-ph/9906538;
            Nucl.~Phys.~B555 (1999) 355, hep-ph/9812413.
\bibitem{AaSm12} 
             G.~Aarts, J.~Smit, 
             Nucl.~Phys.~B511 (1998) 451, hep-ph/9707342;
              Phys.~Lett.~B393 (1997) 395, hep-ph/9610415.
\bibitem{Ar97}
                P.~Arnold, Phys.~Rev.~D55 (1997) 7781, hep-ph/9701393;
\bibitem{Nautaea} 
                B.-J.~Nauta,
                Nucl.~Phys.~B575 (2000) 383, hep-ph/9906389;
               G.~Aarts, B.-J.~Nauta, Ch.~G.~van Weert,
               Phys.~Rev.~D61 (2000) 105002, hep-ph/9911463.
\bibitem{1oNph} D.~Boyanovsky, H.J.~de Vega, astro-ph/0006446.
\bibitem{1oNDCC} 
               F.~Cooper, Y.~Kluger, E.~Mottola, J.P.~Paz,
               Phys.~Rev.~D51 (1995) 2377;
               M.A.~Lampert, J.~Dawson, F.~Cooper,
               Phys.Rev. D54 (1996) 2213, hep-th/9603068.
\bibitem{MiCoDa97}
               B.~Mihaila, F.~Cooper and J.F.~Dawson,
               Phys.~Rev.~D56 (1997) 5400, hep-ph/9705354.
\bibitem{Miea00} 
              B.~Mihaila, T.~Athan, F.~Cooper, J.~Dawson, S.~Habib,
              Phys.Rev. D62 (2000) 125015, hep-ph/0003105.
\bibitem{RyYa00} A.V.~Ryzhov and L.G.~Yaffe, 
                 Phys.~Rev.~D62 (200) 125003, hep-ph/0006333.
\bibitem{MiCoDa00} 
              B.~Mihaila, F.~Cooper and J.F.~Dawson, 
              Phys.~Rev.~D63 (2001) 046003, hep-ph/0006254.
\bibitem{BeCo00} J.~Berges and J.~Cox, hep-ph/0006160.
\bibitem{Boyea96}
              D.~Boyanovsky, H.~J.~de Vega, R.~Holman, J.~F.~J.~Salgado,
              Phys.~Rev.~D54 (1996) 7570, hep-ph/9608205.
\bibitem{MaWo91} L.~Mandel and E.~Wolf, 
                 {\em Optical coherence in quantum optics}, 
                 (Cambridge University Press, Cambridge 1995);
                 J.R.~Klauder and B.-S. Skagerstam, 
                 {\em Coherent States}, World Scientific 1985.
\bibitem{RoMa96} H.-S.~Roh, T.~Matsui, 
                 Eur.~Phys.~J.~A1 (1998) 205, nucl-th/9611050.
\bibitem{DeWi64} B.S.~DeWitt,
                 {\em Dynamical theory of groups and fields},
                 (Blackie 1965,
                 in `Relativity, Groups and Topology',
                 Gordon and Breach 1964). 
\bibitem{Cooea97}
                 F.~Cooper, S.~Habib, Y.~Kluger, E.~Mottola,
                 Phys.~Rev.~D55 (1997) 6471, hep-ph/9610345.
\bibitem{SaSmVi00ab}
                 M.~Sall\'e, J.~Smit and J.C.~Vink, 
                 hep-ph/0008122;
                 Nucl.~Phys.~(Proc.~Suppl.) 94 (2001) 427, hep-lat/0010054. 
\bibitem{SaSmVi00c} M.~Sall\'e, J.~Smit and J.C.~Vink, hep-ph/0012362.
\end{references}
\end{document}